\documentclass[11pt,urlcolor=blue, linkcolor=blue]{article} 
\usepackage{cite}

\usepackage{hhline}
\usepackage{amsmath, amsthm, amssymb,slashed}
\usepackage{ifpdf}
\ifpdf
  \usepackage[pdftex]{graphicx}
  \usepackage{epstopdf}
\else
  \usepackage[dvips]{graphicx}
\fi
\parskip=\baselineskip

\usepackage{slashed}
\usepackage[makeroom]{cancel}
\usepackage[normalem]{ulem}
\usepackage{soul}
\newcommand{\stkout}[1]{\ifmmode\text{\sout{\ensuremath{#1}}}\else\sout{#1}\fi}
\usepackage{color}
\usepackage[colorlinks,citecolor=blue]{hyperref}

\usepackage{datetime}

\allowdisplaybreaks[1]

\newcommand{\ket}[1]{|#1\rangle}

\newcommand{\innerproduct}[2]{\langle #1| #2\rangle}
\newcommand{\outerproduct}[2]{|#1\rangle\langle #2|}
\newcommand{\ztzt}{\mathbb{Z}_2 \times \mathbb{Z}_2}
\newcommand{\ztwo}{\mathbb{Z}_2}

\newcommand{\hgc}{H^2(G,U(1))}
\newcommand{\hgcinput}[1]{H^2(#1,U(1))}
\newcommand{\hdc}{H^{d+1}(G,U(1))}
\newcommand{\hdcinput}[1]{H^{d+1}(#1,U(1))}
\newcommand{\pf}{{P}_f}
\newcommand{\ztwof}{\mathcal{Z}_2^f}

\usepackage{upgreek}

\usepackage{tabularx}

\newcommand{\ccblue}[1]{\textcolor{blue}{#1}}

\DeclareMathAlphabet{\mathpzc}{OT1}{pzc}{m}{it}

\usepackage{array}
\newcolumntype{M}[1]{>{\centering\arraybackslash}m{#1}}

\newcommand{\bea}{\begin{eqnarray}}
\newcommand{\eea}{\end{eqnarray}}
\def\be{\begin{equation}}
\def\eeq{\end{equation}}

\def\ket#1{\left|#1\right\rangle}

\definecolor{red}{rgb}{1,0,0}
\definecolor{blue}{rgb}{0,0,1}
\definecolor{dblue}{rgb}{0,0,0.4}
\definecolor{green}{rgb}{0,1,0}
\definecolor{black}{rgb}{0,0,0}
\definecolor{white}{rgb}{1,1,1}

\definecolor{brn}{rgb}{.8,.4,.0}
\definecolor{redo}{rgb}{1,.5,.0}
\definecolor{ddgrn}{rgb}{0,0.4,0}
\definecolor{dgrn}{rgb}{0,0.55,0}
\definecolor{dbl}{rgb}{0,0,0.5}

\usepackage[bbgreekl]{mathbbol}
\usepackage{amscd}

\newcommand{\Z}{\mathbb{Z}}

\newcommand{\bpm}{\begin{pmatrix}}
\newcommand{\epm}{\end{pmatrix}}
\newcommand{\bmm}{\begin{matrix}}
\newcommand{\emm}{\end{matrix}}

\newcommand{\cT}{ {\cal T} } 
 
\newcommand{\cZ}{ {\cal Z} } 

\usepackage[all,cmtip]{xy}
\usepackage{tikz-cd}
\usepackage{tikz}
\usetikzlibrary{matrix}

\def\CU{{\cal U}}

\def\CZ{{\cal Z}}

\def\Z{{\mathbb{Z}}}

\begin{document}
\renewcommand{\arraystretch}{1.52}
\begin{titlepage}

\begin{flushright}
\end{flushright}
\vskip 1.25in
\begin{center}

{\bf\LARGE{ 
Unwinding Short-Range Entanglement
\\[3mm]
}}

\vskip.8cm 
\Large{Abhishodh Prakash$^{1,2}$},
\Large{Juven Wang$^3$}, 
\Large{Tzu-Chieh Wei$^{1,2}$}
\vskip.5cm
 {\small{\textit{$^1$Department of Physics and Astronomy, State University of New York at Stony Brook,\\[.4cm] 
 Stony Brook, NY 11794-3840, USA}\\}}
\vskip.4cm
 {\small{\textit{$^2$ 
 C. N. Yang Institute for Theoretical Physics, Stony Brook, NY 11794-3840, USA}\\}}
\vskip.4cm
 {\small{\textit{$^3$School of Natural Sciences, Institute for Advanced Study, Princeton, NJ 08540, USA}\\}}
\end{center}

\vskip.5cm
\baselineskip 16pt
\begin{abstract}

Symmetry-Protected Topological (SPT) phases are gapped phases of quantum matter protected by global symmetries 
that cannot be adiabatically deformed to a trivial phase without breaking symmetry. In this work, we show that, for several SPT phases that are short range entangled (SRE), enlarging symmetries may effectively achieve the consequences of explicitly breaking symmetries. In other words, we demonstrate that non-trivial SPT phases can be unwound to trivial ones by \emph{symmetry extension} --- 
through a path where the Hilbert space is \emph{enlarged} and
the Hamiltonian is invariant under an \emph{extended} symmetry group applying the idea of Wang, Wen and Witten in arXiv:1705.06728. We show examples of both bosonic and fermionic SPT phases in 1+1 dimensions,
including Haldane's bosonic spin chain and layers of Kitaev's fermionic Majorana chains. By adding degrees of freedom into the boundary/bulk, 
we can lift the zero mode degeneracy, or unwind the whole system. Furthermore, based on properties of Schur cover, we sketch a general picture of unwinding applicable to any 1+1 D bosonic SPT phase protected by on-site finite symmetry. Altogether we show that SRE states can be unwound by \emph{symmetry breaking}, \emph{inversion} and \emph{symmetry extension}. 

\end{abstract}
 
\end{titlepage}

\tableofcontents   


\section{Introduction and summary of main results} \label{sec:intro}

\begin{figure}[!htbp]
	\centering	
	\includegraphics[width=100mm]{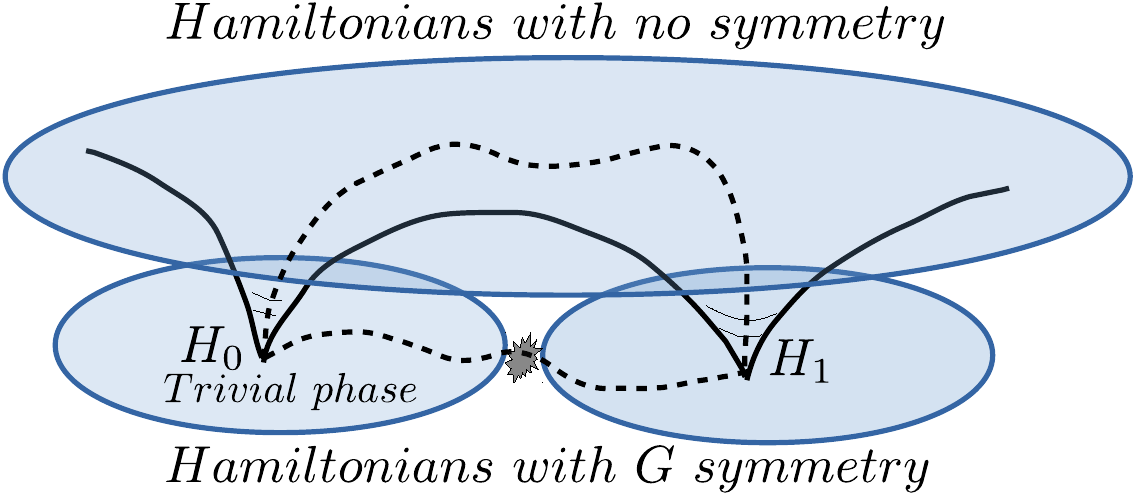}
	\caption{$H_1$, which belongs to the trivial phase in the space of Hamiltonians without symmetry, can become non-trivial in the space of Hamiltonians with some symmetry $G$ .\label{fig:ham_breaking}}
\end{figure}

Gapped phases of quantum matter can be thought of as equivalence classes of physical systems, whose dynamics are governed by local Hamiltonians with a spectral gap. Two gapped Hamiltonians are said to be equivalent, i.e., the physical systems described by them belong to the same phase if they can be interpolated without closing the spectral gap. The presence of global symmetries, which is natural in many condensed matter systems adds an additional degree of complexity and results in an increase in the number of possible fine-grained phases. A Hamiltonian that belongs to the trivial phase within the space of gapped Hamiltonians without any symmetry constraint may become non-trivial in the space of \emph{symmetric} gapped Hamiltonians as shown in Fig.~\ref{fig:ham_breaking}. One well known mechanism by which phases can appear due to the presence of symmetries  is when the global symmetry is \emph{spontaneously broken}  \'a la Ginzburg and Landau. Interestingly, even when symmetry is unbroken, it was recently discovered that we can have different phases that cannot be connected to each other without a phase transition. Such phases are called symmetry-protected-topological (SPT) phases, which are the focus of our current study. 

There has been a great deal of interest in recent years in characterizing and classifying SPT phases in various spatial dimensions. This is in part due to the successful prediction and experimental detection of topological insulators and in part due to the rich theoretical structure that has been uncovered in understanding these phases (see Refs. \cite{HasanKane_2010RevModPhys_TopInsulators,QiZhang_RevModPhys2011_TopIns,KaiTeoSchnyderRyu_RevModPhys2016_ClassificationOfTopWithSymm,Ludwig_2016PhysicaScripta_ClassificationReview, 1610.03911Wen} for reviews). Let us review some important facts about non-trivial SPT phases with a global symmetry $G$:

 \emph{Fact 1: The ground state of any Hamiltonian describing a non-trivial SPT phase cannot be mapped to a trivial state (e.g., product state for bosons, slater determinant state for fermions) using a finite-depth unitary circuit (FDUC) with each layer being invariant under $G$.} 
\begin{figure}[!htbp]
	\centering	
	\includegraphics[width=100mm]{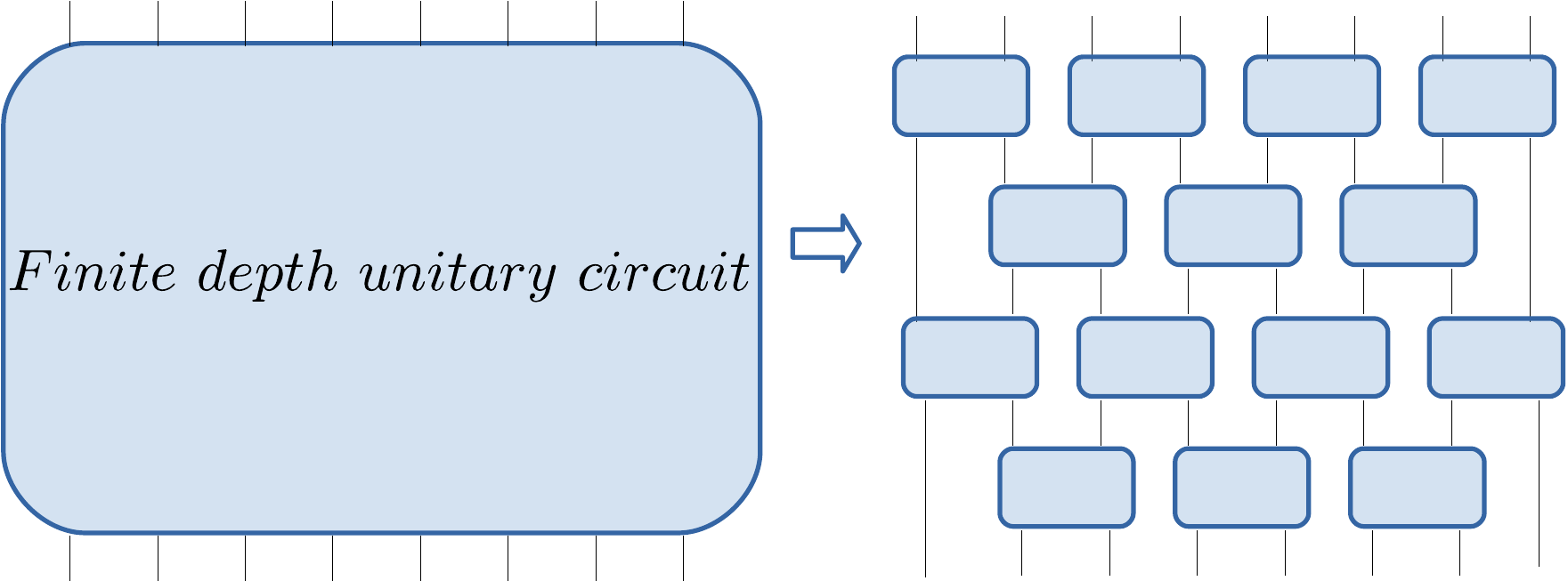}
	\caption{A finite depth unitary circuit (FDUC).\label{fig:fduc}}
\end{figure}
 
An FDUC is a unitary operator that can be written as the product of a finite number of ultra-local unitary operators of the form $\bigotimes_i u_i$ where each $u_i$ operates on a disjoint Hilbert space associated to a finite number of lattice points close to the site $i$ as shown in Fig.~\ref{fig:fduc}. It is easy to see that any FDUC can only produce short-range entanglement. Fact 1 is an alternative way of phrasing the fact that the Hamiltonian cannot be connected to a trivial one via a path of gapped Hamiltonians that are invariant under $G$. We can ask important questions about the precise conditions under which a non-trivial SPT phase can or cannot be unwound to a trivial one. For instance,
 
 \emph{Q1: How much symmetry needs to be broken to be able to map the ground state of a non-trivial SPT phase to a product state using an FDUC?} 

\begin{figure}[!htbp]
	\centering
\begin{tabular}{cc}			
	\includegraphics[width=80mm]{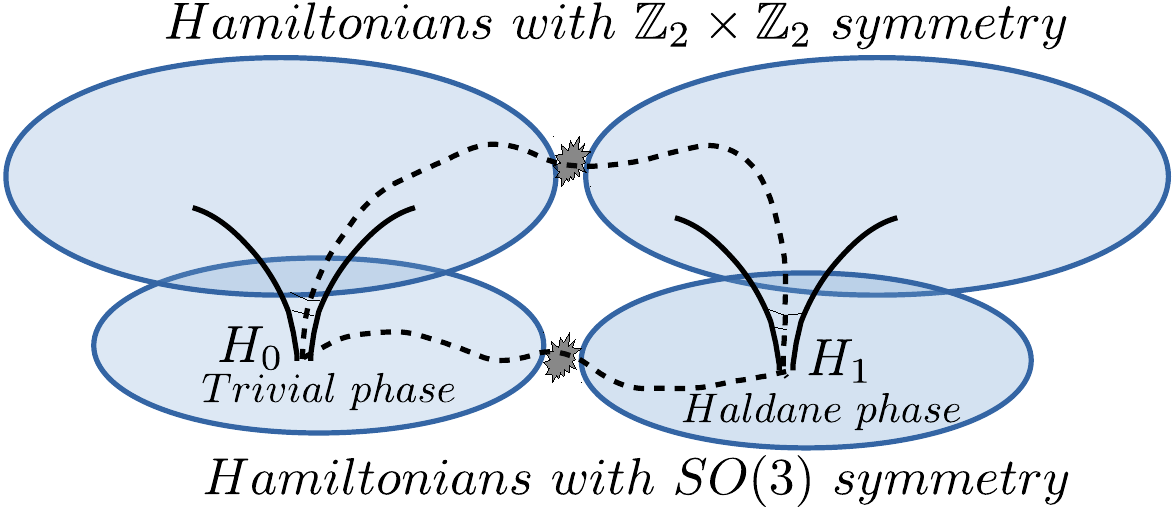} & 	
	\includegraphics[width=80mm]{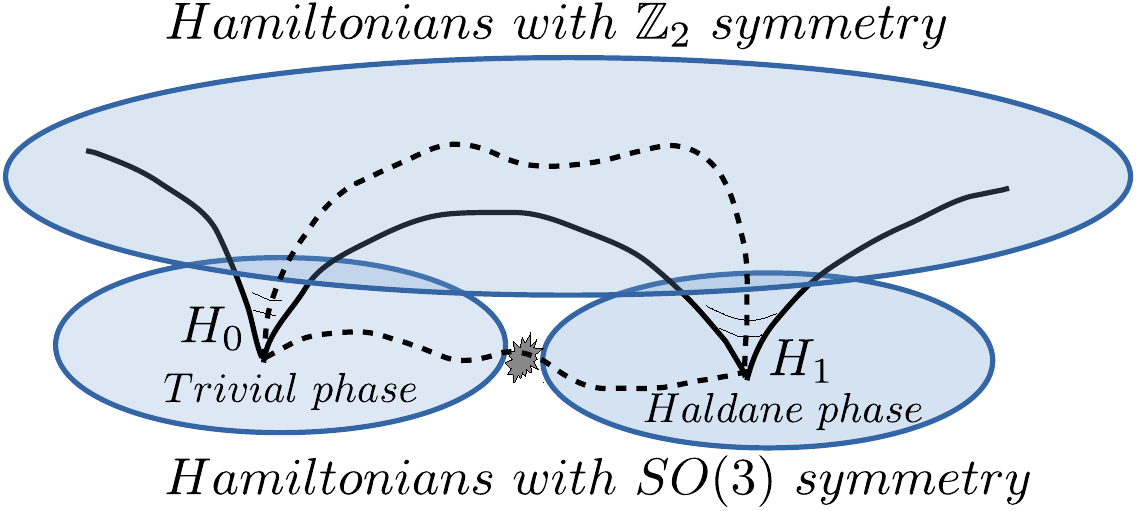} 
\end{tabular}		
	\caption{Unwinding the Haldane phase by explicitly breaking symmetry.  (FDUC).\label{fig:ham_breaking_Haldane}}
\end{figure}

To answer this, let us consider the famous example of the AKLT model~\cite{AKLT_1987PhysRevLett_AKLT}, which is invariant under an on-site action of the group $SO(3)$ and belongs to the so-called Haldane phase. It is known that certain essential features of the Haldane phase, such as the emergent fractionalized boundary modes are present even if $SO(3)$ is explicitly broken down, using weak perturbations, to its abelian subgroup, $\ztzt$ comprising of $\pi$ rotations about the $x$, $y$ and $z$ axes~\cite{PollmannTurnerBergOshikawa_2010PhysRevB_EntanglementSpectrumSPT,PollmannTurnerBergOshikawa_2012PhysRevB_SymmetryProtectionSPT}. However, if the symmetry is broken down further to $\ztwo$ (leaving behind no other accidental symmetries like inversion), generated by $\pi$ rotations only about one of the axes, then the phase becomes trivial! This means that we cannot use a $\ztzt$ invariant path to unwind the AKLT ground state but we can use a $\ztwo$ invariant one as shown in Fig.~\ref{fig:ham_breaking_Haldane}. This above result can be understood within the group-cohomology classification framework which posits that in $d$ spatial dimensions, bosonic SPT phases are classified by the elements of the cohomology group $\hdc$. The 1+1 D AKLT model is non-trivial in the sense that it corresponds to the non-trivial element of $\hgcinput{SO(3)} \cong \ztwo$. Now, upon restricting the group $SO(3)$ to $\ztzt$ by introducing symmetry-breaking perturbations to the AKLT Hamiltonian, it turns out that the system still belongs to a non-trivial SPT phase, now labeled by the non-trivial element of $\hgcinput{\ztzt} \cong \ztwo$. However, since $\hgcinput{\ztwo} \cong 1$, upon further breaking the symmetry down to $\ztwo$, we are only left with the trivial SPT phase. 

Let us phrase the general condition a bit more technically. Given a group $G$, we can specify a subgroup $K$ of $G$ with an injective homomorphism, 
\begin{equation} \label{eq:subgroup}
i: K \rightarrow G.
\end{equation}
An SPT phase protected by $G$ is characterized by a set of \emph{cocycles} $\omega_{d+1}(\{g_i\})$ whose class corresponds to an element of the group $\hdc$ (see Sec.~\ref{sec:bosonic} for more details). The map of Eq.~\ref{eq:subgroup} allows us to define a set of cocycles of $K$ via pullback $i^*\omega_{d+1}(\{k_i\}) = \omega_{d+1}(\{i(k_i)\})$. Using this information, we can give the answer to $Q1$:

  \emph{An SPT phase with global symmetry G classified by a set of cocycles $\omega_{d+1}(\{g_i\})$ whose class corresponds to a non-trivial element of $\hdc$ can be trivialized by breaking $G$ to $K$ related by an injective homomorphism $i:K\rightarrow G$ if the class corresponding to the cocycles of $K$ defined via pullback $i^*\omega_{d+1}(\{k_i\})$ corresponds to the trivial element of $\hdcinput{K}$}. A corollary of this result is that a guaranteed way to trivialize any SPT phase is by breaking all symmetries i.e. $K \cong 1$. 
  
  We now ask a second question which is, in some sense converse to $Q1$: 
 
 \emph{Q2: Instead of breaking the symmetry, can we find a way to unwind an SPT phase by extending the global symmetry?}

The answer to the above question is yes and the theoretical justification is established in Ref.~\cite{WangWenWitten_PhysRevX.8.031048_symmetric} where the authors provide a new perspective on another fact about SPT phases:
 
\emph{Fact 2: The symmetry action on the boundary of a non-trivial SPT phase suffers from an 't Hooft anomaly. This presents an obstruction to gauging the symmetry and also producing a short-range-entangled symmetric gapped Hamiltonian for the boundary degrees of freedom.}

The authors of Ref.~\cite{WangWenWitten_PhysRevX.8.031048_symmetric} show how to systematically produce a symmetric gapped Hamiltonian at the boundary by suitably extending $G$ to  $\tilde{G}$ and dynamically gauging the anomaly-free normal subgroup, $K$ of $\tilde{G}$ by which $G$ was extended. This leaves behind a $\tilde{G}/K \cong G$ symmetric theory as desired. It is important to note that the choice of groups $\tilde{G}$ and $K$ that satisfy the above requirements are not unique and in Ref.~\cite{WangWenWitten_PhysRevX.8.031048_symmetric}, the authors provide examples demonstrating this. The presence of emergent gauge degrees of freedom however renders the boundary long-range entangled which is consistent with the expectation that we cannot have a short-range entangled symmetric boundary for a non-trivial SPT phase. Let us phrase this result a little more technically which will help us answer $Q2$:

\emph{An SPT phase with global symmetry G classified by a set of cocycles $\omega_{d+1}(\{g_i\})$ whose class corresponds to a non-trivial element of $\hdc$ can be trivialized by extending  $G$ to $\tilde{G}$ which are related by a surjective homomorphism, $s:\tilde{G} \rightarrow G$ such that the class of cocycles of $\tilde{G}$ defined by pullback $s^*\omega_{d+1}(\{\tilde{g}_i\})$ corresponds to the trivial element of $\hdcinput{\tilde{G}}$.
}

\begin{figure}[!htbp]
	\centering	
	\includegraphics[width=100mm]{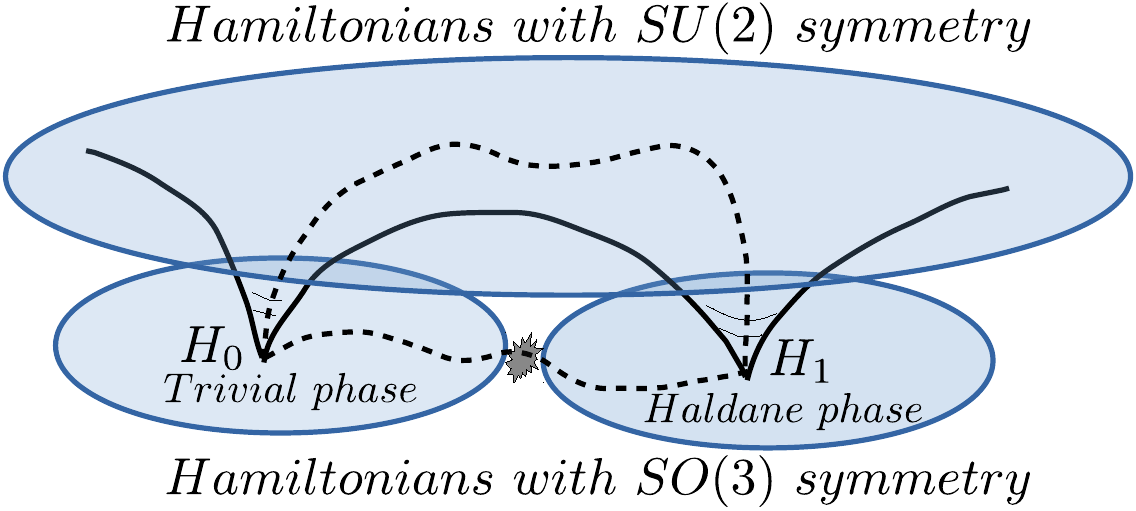}
	\caption{Trivializing the Haldane phase by symmetry extension. \label{fig:extension}}
\end{figure}

 To put this in perspective, let us again consider the $SO(3)$ invariant Haldane phase. A Hamiltonian belonging to this phase like the AKLT model cannot be connected to the trivial phase in the space of $SO(3)$ invariant Hamiltonians. However, they can be connected in the space of the larger $SU(2)$ invariant Hamiltonians as shown in Fig.~\ref{fig:extension}. Here, $SU(2)$ is the required extension to $SO(3)$ as described above. What it physically means to extend symmetry and connect the system to the trivial phase (i.e. unwind the system) is explored in some detail in this paper. 

The main purpose of this paper is to explicitly demonstrate the affirmative answer above to $Q2$ for a large class of SPT phases employing the results of Ref.~\cite{WangWenWitten_PhysRevX.8.031048_symmetric}. The rest of the paper is organized as follows. In Sec.~\ref{sec:bosonic} we discuss unwinding of nontrivial bosonic SPT phases, including representative states in the Haldane phase (interpreted as an SPT phase protected by different symmetries) and the cluster state. We provide a general picture for unwinding nontrivial (1+1)D SPT phases protected by finite on-site symmetry. In Sec.~\ref{sec:fermionic} we turn to unwinding nontrivial fermionic SPT phases. Five of the ten Altland-Zirnbauer symmetry classes in (1+1)D have a non-trivial classification in the free-fermionic limit and some of these are reduced in the presence of interactions. These classes are D, DIII, BDI, AIII and CII.  Representative models of non-trivial SRE phases belonging to all of these classes can be constructed by stacking Kitaev's Majorana chains~\cite{Kitaev_2001PhysicsUspekhi_KitaevChain} (henceforth referred to as the Kitaev chain) and are shown in Appendix.~\ref{app:fermion_hamiltonian_stracking}. In Sec.~\ref{sec:fermionic}, we show that some of these non-trivial fermionic models that can be understood as bosonic SPT phases can be unwound by a suitable symmetry extension. In Sec.~\ref{sec:conclude}, we summarize and make some concluding remarks.

We remark on the notation of symmetry groups. We use the `mathcal' convention for symmetry groups that contains the fermionic parity operator $(-1)^{N_f}$ in the group center.
For example, the group of time reversal symmetry generated by $\cT$ such that $\cT^2 = (-1)^{N_f}$ is denoted as $\mathcal{Z}_4^T = \{1,\cT,(-1)^{Nf}, (-1)^{Nf} \cT \}$. On the other hand, the group of time reversal symmetry generated by $\cT$ such that $\cT^2 = 1$ is denoted as $\ztwo^T = \{1,\cT \}$.

\section{Two known roads to unwinding SPT phases and a third one} \label{sec:three roads}
\begin{figure}[!htbp]
	\centering	
	\includegraphics[width=100mm]{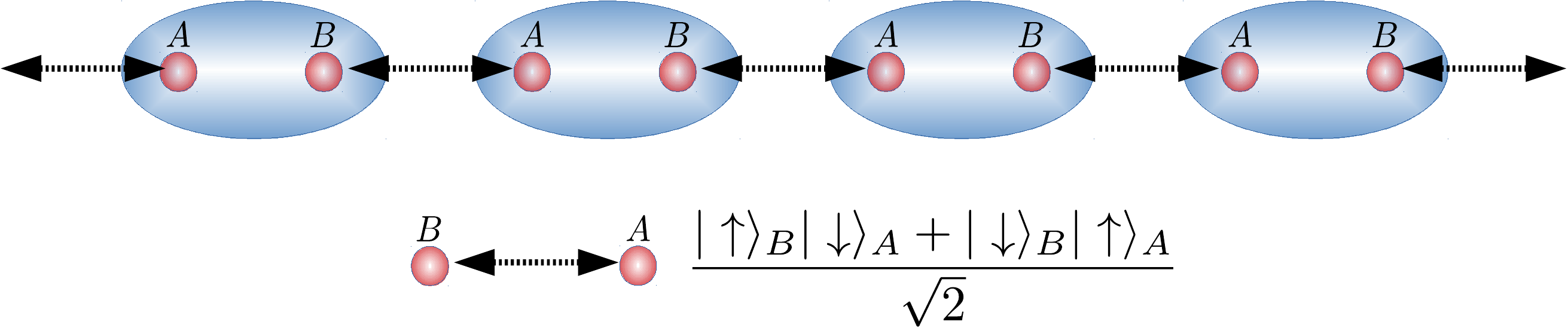}
	\caption{A representative SPT state. \label{fig:caricature}}
\end{figure}

In this section, we review two known ways of mapping a non-trivial SPT state to a trivial one using a FDUC-- symmetry breaking and inversion. We then introduce the third way, symmetry extension which will form the subject matter for the rest of the paper. We use a representative caricature of an SPT state shown in Fig.~\ref{fig:caricature} formed by considering two qubits per unit site and maximally entangling the neighboring qubits on different sites:
\begin{equation}
\ket{\psi} = \prod_{k } \left(  \frac{\ket{\uparrow}_{B,k}\ket{\downarrow}_{A,k+1}+ \ket{\downarrow}_{B,k}\ket{\uparrow}_{A,k+1}}{\sqrt{2}} \right). 
\end{equation}
This state represents a non-trivial SPT ground state protected symmetry group $\ztzt$ generated by the two commuting operators, $\prod_{k} \sigma^x_{A,k} \sigma^x_{B,k}$ and $\prod_{k} i\sigma^z_{A,k} i\sigma^z_{B,k}$ in that it cannot be mapped to a trivial product state using a FDUC where each layer commutes with the symmetry generators.  We will return to this state and also write down its zero correlation length fixed-point Hamiltonian explicitly in Sec.~\ref{sec:bosonic}. We now proceed to trivializing the state. 

\subsection{Explicit symmetry breaking}
\begin{figure}[!htbp]
	\centering		
	\includegraphics[width=100mm]{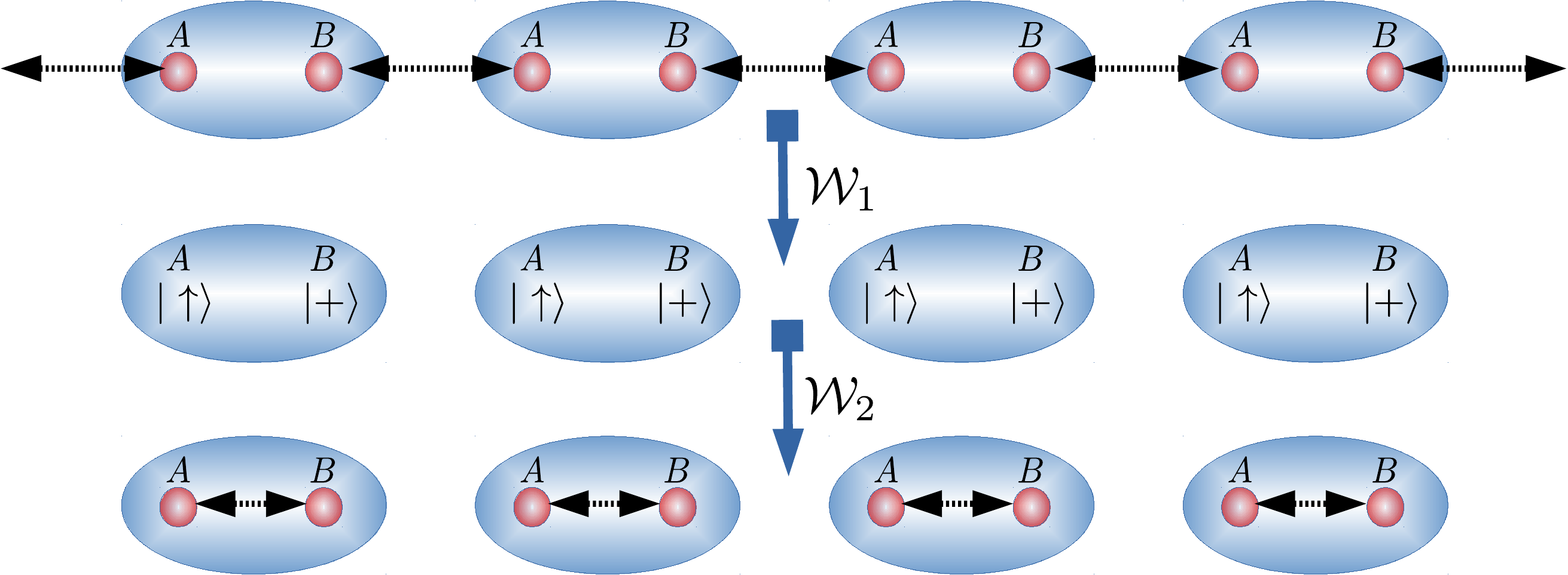}
	\caption{Unwinding by explicit symmetry-breaking. \label{fig:caricature_breaking}}
\end{figure}

Consider the two-layer FDUC, $\mathcal{W} = \mathcal{W}_2 \mathcal{W}_1$
\begin{eqnarray}
\mathcal{W}_1 &=& \prod_k \left[\outerproduct{\uparrow}{\uparrow}_{B,k} \otimes \sigma^x_{A,k+1} + \outerproduct{\downarrow}{\downarrow}_{B,k} \otimes \mathbb{1}_{A,k+1} \right] \\
\mathcal{W}_2 &=& \prod_k \left[\outerproduct{\uparrow}{\uparrow}_{B,k} \otimes \sigma^x_{A,k} + \outerproduct{\downarrow}{\downarrow}_{B,k} \otimes \mathbb{1}_{A,k} \right] 
\end{eqnarray}
Applying $\mathcal{W}$ to $\ket{\psi}$ leaves us with the trivial product state, $\ket{\psi_0}$ as shown in Fig.~\ref{fig:caricature_breaking},
\begin{equation}\label{eq:trivial}
\mathcal{W} \ket{\psi} = \ket{\psi_0} = \prod_k \left(  \frac{\ket{\uparrow}_{A,k}\ket{\downarrow}_{B,k}+ \ket{\downarrow}_{A,k}\ket{\uparrow}_{B,k}}{\sqrt{2}} \right)
\end{equation}
However, $\mathcal{W}_1$ and $\mathcal{W}_2$ do not commute with the symmetry operators $\prod_{k} \sigma^x_{A,k} \sigma^x_{B,k}$ and $\prod_{k} i\sigma^z_{A,k} i\sigma^z_{B,k}$ and hence this is a case of unwinding by explicit symmetry-breaking.

\subsection{Inversion}
SPT phases are said to be invertible, meaning that for every non-trivial SPT phase, we can find its inverse phase, which, if stacked on the original SPT phase can be unwound together to a trivial one. This follows from the fact that SRE phases have an abelian group structure with respect to stacking. If a phase, labeled by an element $\alpha$  is stacked on another phase, labeled by $\beta$, the net system is a phase labeled by $\alpha + \beta$. The non-trivial SPT state we are considering has a  $\ztwo$ classification from group-cohomology (see Sec.~(\ref{sec:bosonic})). This means that the non-trivial phase is its own inverse and by stacking two layers of the system, we should be able to map it to a trivial state using a FDUC that commutes with the  $\ztzt$ symmetry  generators at each layer. Let us check this explicitly. 

First, let us consider the ground state of two stacked SPT phases:
\begin{equation}
\ket{\tilde{\psi}} = \ket{\psi}_1 \otimes \ket{\psi}_2 = \prod_{k } \prod_{\alpha = 1,2} \left(  \frac{\ket{\uparrow}_{B,\alpha,k}\ket{\downarrow}_{A,\alpha,k+1}+ \ket{\downarrow}_{B,\alpha,k}\ket{\uparrow}_{A,\alpha,k+1}}{\sqrt{2}} \right)
\end{equation}

\begin{figure}[!htbp]
	\centering	
	\includegraphics[width=100mm]{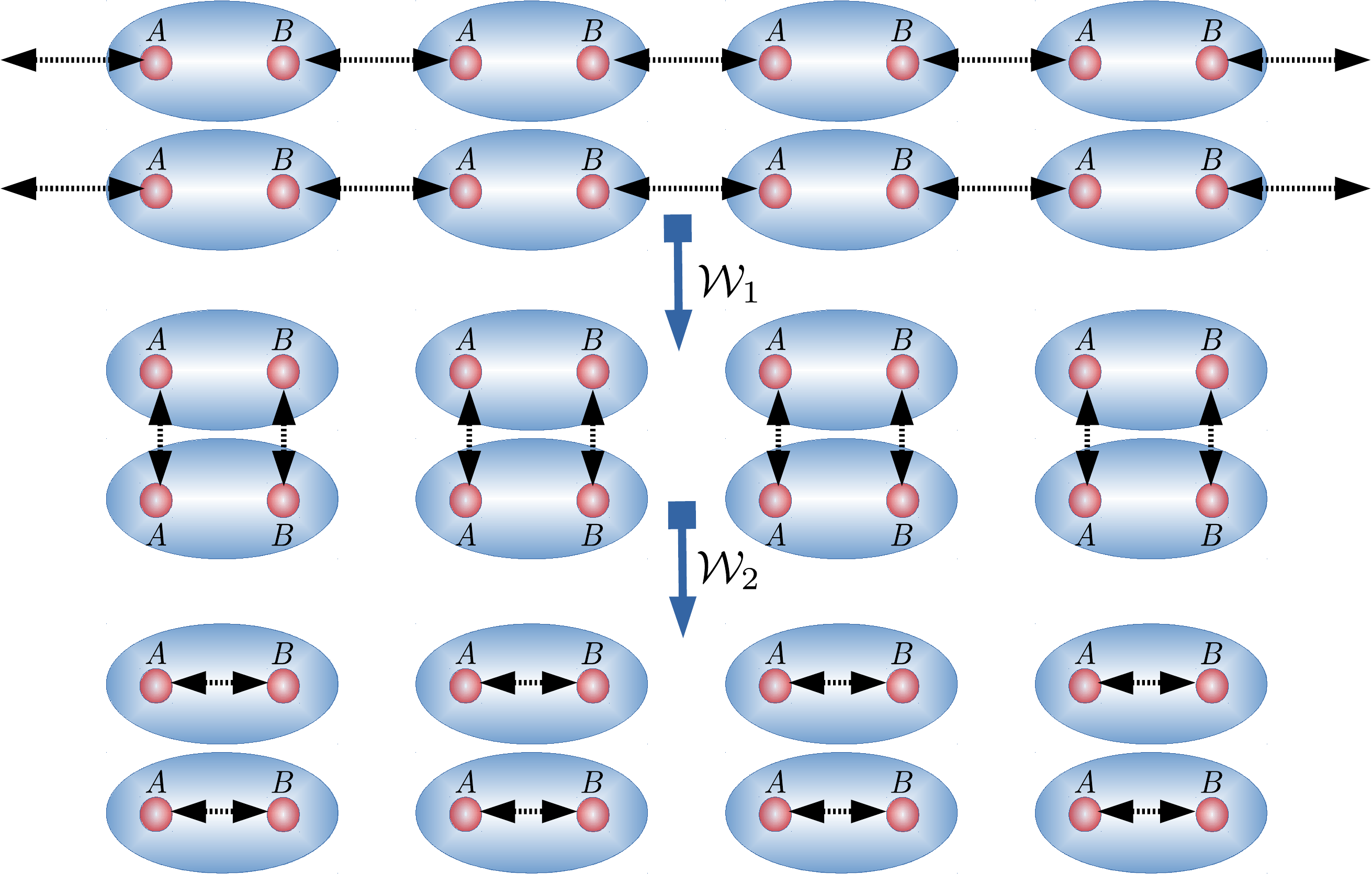}
	\caption{Unwinding by inversion. \label{fig:caricature_inversion}}
\end{figure}

As shown in fig.~\ref{fig:caricature_inversion}, we can use the following two-layer FDUC to map this state to two layers of the trivial state of Eq.~\ref{eq:trivial}. 
\begin{eqnarray}
\mathcal{W}_1 &=& \prod_k \frac{1}{2} \left(\mathbb{1} + \vec{\sigma}_{B,1,k}\cdot \vec{\sigma}_{A,2,k+1}\right) \\
\mathcal{W}_2 &=& \prod_k \frac{1}{2} \left(\mathbb{1} + \vec{\sigma}_{A,1,k}\cdot \vec{\sigma}_{B,2,k}\right) \\
\mathcal{W} \ket{\tilde{\psi}} &=& \prod_{k } \prod_{\alpha = 1,2} \left(  \frac{\ket{\uparrow}_{B,\alpha,k}\ket{\downarrow}_{A,\alpha,k}+ \ket{\downarrow}_{B,\alpha,k}\ket{\uparrow}_{A,\alpha,k}}{\sqrt{2}} \right),
\end{eqnarray}

\noindent where $\mathcal{W} = \mathcal{W}_2 \mathcal{W}_1$. The operator $\frac{1}{2} \left(\mathbb{1} + \vec{\sigma}_{A}\cdot \vec{\sigma}_{B}\right)$ is a swap operator that exchanges the basis states $\ket{\uparrow},~\ket{\downarrow}$ on two sites, $A$ and $B$, and is easily checked to commute with the $\ztzt$ symmetry generators. Thus, we have unwound the SPT phase without breaking symmetry but by stacking an `inverse phase'.

\subsection{Symmetry extension}
\begin{figure}[!htbp]
	\centering	
	\includegraphics[width=100mm]{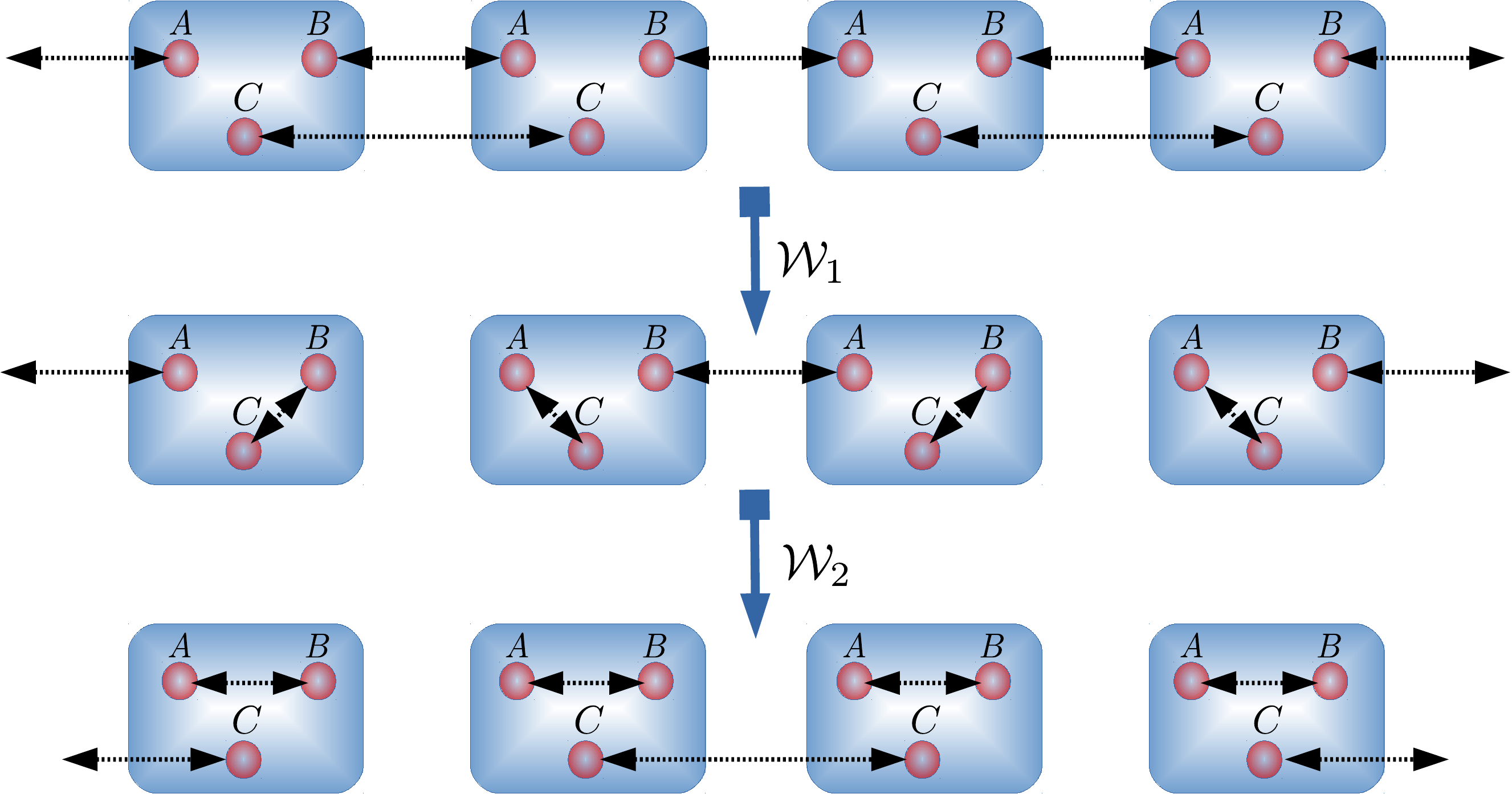}
	\caption{Unwinding by symmetry extension. \label{fig:Cluster_trivialize1}}
\end{figure}
Let us now consider unwinding the SPT state by symmetry extension. In order to do this, we stack a product state of dimers to the original SPT state and increase the local Hilbert space dimension. 
\begin{eqnarray}~\label{eq:extended_state_d8}
\ket{\tilde{\psi}} = \ket{\psi} \otimes \prod_{\text{odd}~k} \frac{(\ket{\downarrow}_{C,k}\ket{\uparrow}_{C,k+1}+\ket{\uparrow}_{C,k}\ket{\downarrow}_{C,k+1})}{\sqrt{2}}
\end{eqnarray}
Note that this is a trivial operation in the sense that we are stacking something that manifestly belongs to the trivial phase. However this helps in increasing the local Hilbert space so that it transforms faithfully under the extended symmetry group generated by the operators  $\prod_{k} \sigma^x_{A,k} \sigma^x_{B,k} \sigma^x_{C,k}$ and $\prod_{k} i\sigma^z_{A,k} i\sigma^z_{B,k} i\sigma^z_{C,k}$ which do not commute with each other. These generators are a faithful representation of the dihedral group of eight elements, $\mathbb{D}_8$ which leaves the state of Eq.~\ref{eq:extended_state_d8} invariant.
As shown in fig.~\ref{fig:Cluster_trivialize1}, this state can be unwound by the application of the following FDUC $\mathcal{W} = \mathcal{W}_2 \mathcal{W}_1$, with each layer $\mathcal{W}_i$ being invariant under the $\mathbb{D}_8$ symmetry,
\begin{eqnarray}
\mathcal{W}_1 &=& \prod_{\text{odd~} k} \frac{1}{2} \left(\mathbb{1} + \vec{\sigma}_{C,k}. \vec{\sigma}_{A,k+1}\right)\\
\mathcal{W}_2 &=& \prod_{\text{odd~} k} \frac{1}{2} \left(\mathbb{1} + \vec{\sigma}_{C,k}. \vec{\sigma}_{A,k}\right) \prod_{\text{even~} k} \frac{1}{2} \left(\mathbb{1} + \vec{\sigma}_{C,k}. \vec{\sigma}_{B,k}\right) \\
\mathcal{W} \ket{\tilde{\psi}} &=& \prod_{k } \left(  \frac{\ket{\uparrow}_{B,k}\ket{\downarrow}_{A,k+1}+ \ket{\downarrow}_{B,k}\ket{\uparrow}_{A,k+1}}{\sqrt{2}} \right)  \prod_{\text{even}~k} \left(\frac{\ket{\downarrow}_{C,k}\ket{\uparrow}_{C,k+1}+\ket{\uparrow}_{C,k}\ket{\downarrow}_{C,k+1}}{\sqrt{2}} \right)
\end{eqnarray}
This is an example of unwinding by symmetry extension which we will explore further. The relationship between the original symmetry group, $\ztzt$ to the extended one, $\mathbb{D}_8$ as well as a number of other details and generalities will be made clear in the following sections.

\section{Unwinding bosonic SPT phases}
\label{sec:bosonic}
In this section, we demonstrate how fixed-point bosonic SPT states and their parent Hamiltonians can be trivialized by symmetry extension. We begin with a short review of the group cohomology classification of bosonic SPT phases, first  in 1+1 D and then in general dimensions. We further review key results from the paper by Wang, Wen and Witten in~\cite{WangWenWitten_PhysRevX.8.031048_symmetric} beyond the details provided in the introduction.  We then demonstrate our trivialization procedure for 1+1 D bulk using the same symmetry-extension procedure on a few specific examples of well-known bosonic SPT phases, and  we also state a general picture for the case of arbitrary on-site finite unitary symmetry. Note that everywhere in this paper, unless stated otherwise, we consider one-dimensional systems of length $L$ assumed to be in the thermodynamic limit ($L>>1$) with lattice constant set to 1 and employ periodic boundary conditions (unless stated otherwise).

\subsection{A quick recap of the classification of bosonic SPT phases in 1+1D and beyond}

We start with a quick recap of the classification of bosonic SPT phases in (1+1)D following Ref~\cite{ChenGuWen_2011PhysRevB_CompleteClassification1d}. Let us first recall that SPT phases are gapped phases of matter with a unique ground state. In (1+1)D, this allows us to represent any such ground state faithfully as a matrix product state (MPS) with a sufficiently large but finite bond dimension $\chi$ that does not scale with the system size~\cite{VerstraeteCirac_2006PhysRevB_MPS,SchuchWolfVerstraete_2008PhysRevLett_MPS}. Let us focus on a spin chain with an on-site Hilbert space of dimension $J$ and choose some basis appropriately labeled $\ket{i} = \ket{1}, \ket{2}, \ldots, \ket{J}$. For convenience of notation, let us also assume lattice translation invariance. An MPS representation of a gapped ground state of such a system can be written using $J$ matrices of size $\chi \times \chi$, $A_1, \ldots, A_J$ as follows
\begin{equation}
\ket{\psi} = \sum_{i_1 = 1}^J \ldots \sum_{i_L = 1}^{J}  Tr\left[A_{i_1} A_{i_2} \ldots A_{i_L} \right] \ket{i_1 \ldots i_L}.
\end{equation}  
First, note that changing $A_i \mapsto M A_i M^{\dagger}$ with any unitary $M$ leaves $\ket{\psi}$ invariant and hence is a redundancy in the MPS representation.
Let us now consider $\ket{\psi}$, a unique ground state, which invariant under the group of symmetry operations, $g \in G$ of Hamiltonian, $g: \ket{\psi} \mapsto \ket{\psi}$.
We can re-express the invariance condition of $\ket{\psi}$ as a condition on the set of matrices $A_i$. The different inequivalent ways of this symmetry action on the matrices $A_i$ effectively give us a classification of different SPT phases. Let us demonstrate this using a few examples starting with time reversal symmetry. 

Consider the action of time-reversal symmetry with an anti-unitary representation, $\mathcal{T}$ such that $\mathcal{T}^2 = 1$. Any time-reversal symmetry operator can be written using an on-site unitary operator, $U(\mathcal{T})$ combined with complex-conjugation, $\mathcal{K}$,
\begin{equation}
\mathcal{T} = \left[\bigotimes_{i=1}^L U(\mathcal{T})\right] ~\mathcal{K}.
\end{equation}
The invariance condition $\mathcal{T} \ket{\psi} = \ket{\psi}$ can translated to the matrices $A_k$ as follows
\begin{equation}
\sum_{k=1}^J U(\mathcal{T})_{ik} A^*_k = V A_i V^\dagger.
\end{equation}
The condition $\mathcal{T}^2 = 1$ imposes the condition $V^* V = \pm \mathbb{1}$ and thus divides the \emph{virtual space} (sometimes also called the \emph{bond space}) symmetry representation $V$ into two classes labeled by $\pm$. This gives us the $\ztwo$ classification of $\mathcal{T}$ invariant spin chains. 

Let us now consider the case of internal unitary symmetries, which is described by an on-site unitary representation of the elements of some group $G$, $\bigotimes_{i=1}^L U(g)$. The invariance condition, $ \bigotimes_{i=1}^L U(g) \ket{\psi} = \ket{\psi}$, can be translated to the level of $A_k$ matrices as follows
\begin{equation} \label{eq:Internal symmetry MPS}
\sum_{k=1}^{J} U(g)_{ik} A_k = V(g) A_i V^\dagger(g).
\end{equation}
Firstly, note that re-phasing the representation of $G$ on the virtual dimension $V(g)$ by a 1D representation, $\beta_1(g)$ as follows is a gauge freedom that leaves  Eq.~\ref{eq:Internal symmetry MPS} invariant
\begin{equation}~\label{eq:Projective representation rephasing}
V(g) \mapsto \beta_1(g) V(g).
\end{equation}
Group theoretic constraints on $U(g)$ further impose conditions on $V(g)$. The composition rule $U(g) U(h) = U(gh)$ requires $V(g)$ only closes up to a U(1) factor
\begin{equation}
V(g) V(h) = \omega_2(g,h) V(gh),
\end{equation}
where $\omega_2(g,h)$ is a $U(1)$ phase factor dependent on $g$ and $h$.
This means that $V(g)$ are \emph{projective representations} of $G$. Furthermore, associativity imposes the following \emph{cocycle} constraint on the phases $\omega_2$:
\begin{equation} \label{eq:Cocycle condition}
 \omega_2(g,h)\omega_2(gh,l)\omega^{-1}_2(g,hl)\omega^{-1}_2(h,l) \equiv (\delta\omega_2)(g,h,l) = \mathbb{1}.
\end{equation}
Equation ~\ref{eq:Projective representation rephasing} defines the following $coboundary$ equivalence relation:
\begin{equation} \label{eq:Coboundary condition}
\omega_2(g,h) \sim \omega_2(g,h) \beta_1(g)\beta_1(h) \beta^{-1}_1(gh)  \equiv \omega_2(g,h) (\delta \beta_1)(g,h).
\end{equation}
The different SPT phases in 1+1 D with symmetry group $G$ are classified by the different equivalence classes of $\omega_2$ with the equivalence relation of Eq.~\ref{eq:Coboundary condition} subject to the condition of Eq.~\ref{eq:Cocycle condition}. These classes are labeled by the elements of the second cohomology group of $G$ with $U(1)$ coefficients, $\hgc$.

A natural generalization of the $\hgc$ classification of bosonic SPT phases in 1+1 dimensions to $d$+1 dimensions is replacing $\hgc$ by $\hdc$~\cite{ChenGuXinWen_2013PhysRevB_SPTClassificationGroupCohomology} which labels equivalence classes of $d+1$ cocycles, $\omega_{d+1}(\{g_i\})$ subject to generalizations of Eqs.~\ref{eq:Cocycle condition} and \ref{eq:Coboundary condition}. This classification is known to capture a large class of bosonic SPT phases although exceptions are known to exist~\cite{VishwanathSenthil_2013PhysRevX_BeyondCohomology,WangSenthil_2013PhysRevB_BeyondCohomology,BurnellChenFidkowskiVishwanath_2014PhysRevB_BeyondCohomology}. One important feature of bosonic SPT phases classified by group cohomology is the presence of an 't Hooft anomaly on the boundary~\cite{Wen_2013_PhysRevDSPTAnomaly,KapustinThorngren_2014PhysRevLett_SPTAnomalies} which has several consequences. First, it presents an obstruction to gauging the symmetry on the boundary by forcing it to have a non-on-site representation~\cite{ElseNayak_2014PhysRevB_SPTAnomaly}. Second, it forbids the boundary from being symmetric, gapped and short-range-entangled (see Ref~\cite{Witten_RevModPhys_Fermionpathintegral} for a nice proof by contradiction). However, it has been known that the boundary can be gapped by breaking symmetry (spontaneously or explicitly), or, more interestingly, accompanied by surface topological order with long-range-entanglement~\cite{FidkowskiChenVishwanath_2013PhysRevX_SurfaceTopologicalOrder,WangPotterSenthil_2013PhysRevB_SurfaceTopologicalOrder,BondersonNayakQi_2013JournalStatMech_SurfaceTopologicalOrder,ChenBurnellVishwanath_2015PhysRevX_SurfaceTopologicalOrder,MetlitskiKaneFisher_2015PhysRevB_SurfaceTopologicalOrder,SeibergWitten_2016PTEP_SurfaceTopologicalOrder}. Reference \cite{WangWenWitten_PhysRevX.8.031048_symmetric} puts the latter route to gapping symmetric boundary for bosonic phases classified by group cohomology in a systematic footing by \emph{symmetry extension} which we briefly review below.

Consider a bosonic SPT phase with a boundary 't Hooft anomaly classified by a $(d+1)$ cocycle $\omega_{d+1}(\{g_i\})$ belonging to a non-trivial class of $\hdc$ meaning $\omega_{d+1}(\{g_i\}) \neq 
\delta\beta_d(\{g_i\})$. It was shown in Ref~\cite{WangWenWitten_PhysRevX.8.031048_symmetric} that given the above data, there exists a group extension $\tilde{G}$ which fits into the following \emph{short exact sequence}.

\begin{equation}~\label{eq:Short exact sequence}
1  \longrightarrow {K}  \overset{i}{\longrightarrow} \tilde{G} \overset{s}{\longrightarrow}  G \longrightarrow 1. 
\end{equation}

As usual, $i$ is an injective map and $s$ is a surjective map (see Ref~\cite{Moore_2014Notes_quantumsymmetries} for an introduction to short exact sequences and group extensions). The short exact sequence is such that if we consider the cocycle for the bigger group, $\tilde{G}$, as defined via \emph{pullback} of the surjective map $s$, then it belongs to the trivial class:
\begin{equation}
\omega_{d+1}(\{\tilde{g}_i\}) = s^*\omega_{d+1}(\{\tilde{g}_i\}) = \omega_{d+1}(\{s(\tilde{g_i})\}) = \delta\beta_d(\{\tilde{g}_i\}).
\end{equation}
This fact was used in Ref~\cite{WangWenWitten_PhysRevX.8.031048_symmetric} to produce gapped boundaries by considering a $\tilde{G}$ invariant boundary theory but with the extra symmetry $K$, being dynamically gauged, leaving the true global symmetry to be $\tilde{G}/K \cong G$. Note that the choice of groups $\tilde{G}$ and $K$ that satisfy the above conditions is not unique but Ref~\cite{WangWenWitten_PhysRevX.8.031048_symmetric} argues that atleast one such choice always exists.

Another consequence of the above result, which is the focus of this paper, is that if we extend the symmetry $G$ to $\tilde{G}$ as prescribed by the short exact sequence~(\ref{eq:Short exact sequence}), we can unwind the non-trivial $G$ SPT to a trivial one in a $\tilde{G}$ invariant path in Hamiltonian space. The rest of the paper is concerned with demonstrating this by constructing a $\tilde{G}$ invariant FDUC to map a non-trivial $G$ SPT state to a trivial one for various symmetries. For each case, we state the extension used and demonstrate unwinding but do not explain how the extension is arrived at. We relegate the reader to Ref~\cite{WangWenWitten_PhysRevX.8.031048_symmetric} for those technical details.

\subsection{Unwinding an AKLT-like spin chain}
\begin{figure}[!htbp]
	\centering	
	\includegraphics[width=100mm]{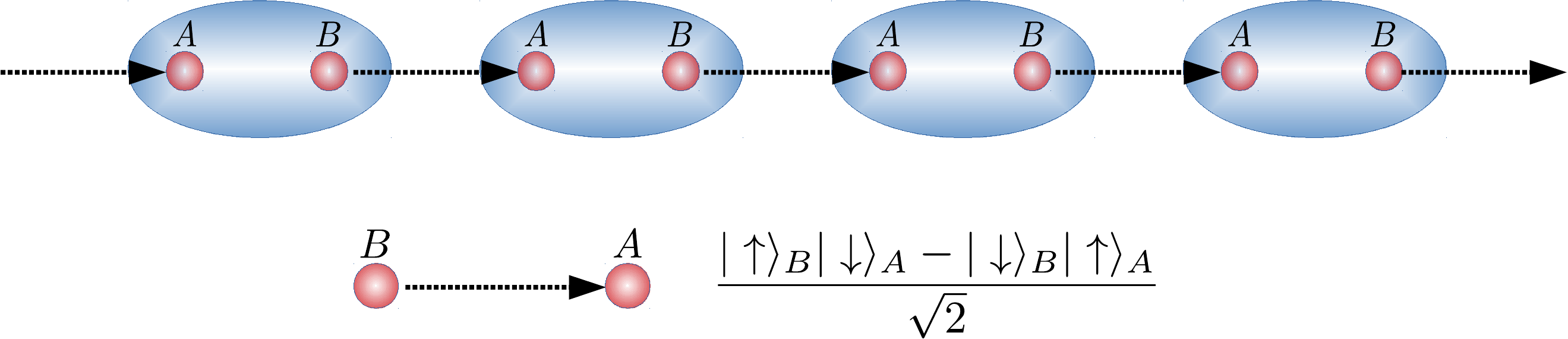}
	\caption{The AKLT-like model. \label{fig:AKLT}}
\end{figure}

The Affleck-Kennedy-Lieb-Tasaki model~\cite{AKLT_1987PhysRevLett_AKLT} is a chain of spin-1 particles with the following Hamiltonian
\begin{equation}
H_{AKLT} = \sum_{j }  \left[\vec{S}_j \cdot\vec{S}_{j+1} + \frac{1}{3} \left( \vec{S}_j \cdot\vec{S}_{j+1}\right)^2\right],
\end{equation}
where $S^\alpha$ are the spin-1 generators of the $SU(2)$ algebra. This Hamiltonian has a unique MPS ground state, which can be written in the basis of the $S^z$ operator, $\ket{+1}, \ket{-1}, \ket{0}$ as follows

\begin{eqnarray}~\label{eq:AKLTstate}
\ket{\psi} &=& \sum_{i_1 = \pm1,0} \ldots \sum_{i_L = \pm1,0}  Tr\left[M_{i_1} M_{i_2} \ldots M_{i_L} \right] \ket{i_1 \ldots i_L}. \\
M_{\pm 1} &=& \pm \sqrt{\frac{2}{3}} \left(\frac{\sigma^x \pm i \sigma^y}{2}\right),~~ M_0 = \frac{-1}{\sqrt{3}} \sigma^z \nonumber.
\end{eqnarray}

 This ground state can also be interpreted as a valence-bond-solid state by first starting with two spin-$\frac{1}{2}$'s per unit site, entangling neighboring spins to form $SU(2)$ singlets and then projecting each site onto the spin-1 sector of the Clebsch-Gordan decomposition $\frac{1}{2} \otimes \frac{1}{2} \cong 1 \oplus 0$.

We now consider a simplified version of the AKLT model shown in Fig.\ref{fig:AKLT}, whose ground state, $\ket{G}$ is the same as $\ket{\psi}$ of Eq.~\ref{eq:AKLTstate} except for the projection onto the spin-1 sector on each site. This leaves us with a 4 dimensional local Hilbert space coming from the two spin halves, which we will call $A$ and $B$, that still transforms as a faithful but reducible $1 \oplus 0$ representation of $SO(3)$. We can also write a parent commuting-projector Hamiltonian $H$, that has $\ket{G}$ as its unique ground state:
\begin{eqnarray} \label{eq:AKLT like}
\ket{G} &=& \prod_{k }  \ket{\psi}_{Bk,Ak+1} = \prod_{k }  \frac{(\ket{\uparrow}_{B,k}\ket{\downarrow}_{A,k+1}- \ket{\downarrow}_{B,k}\ket{\uparrow}_{A,k+1})}{\sqrt{2}}, \\
H &=& - \sum_{k }  \outerproduct{\psi}{\psi}_{Bk,Ak+1}.
\end{eqnarray}

This model has all the appealing features of the AKLT model like fractionalized boundary spins, unique ground state with periodic boundary conditions and a spectral gap, with the added advantage of being exactly solvable. We now unwind this model by interpreting it as two different non-trivial SPT phases protected by two different global symmetries.

\subsubsection{As an $SO(3)$-invariant SPT phase}
\begin{figure}[!htbp]
	\centering	
	\includegraphics[width=100mm]{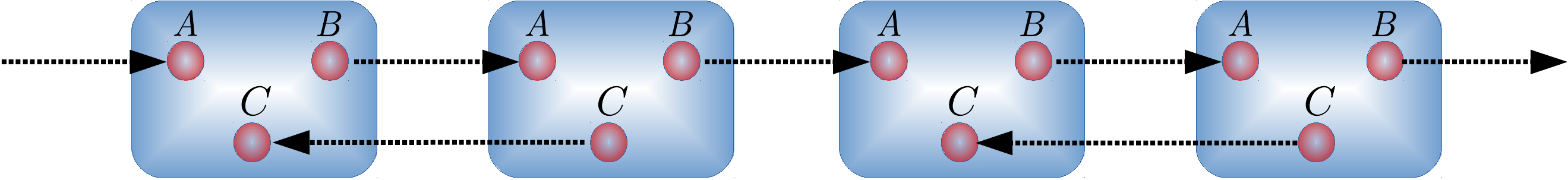}
	\caption{The AKLT-like model with extension. \label{fig:AKLT_extended}}
\end{figure}

If we disregard all other symmetries  except for $SO(3)$ with the following on-site unitary representation
\begin{equation}
U(\theta) = \prod_{k }  \exp\left[ i \theta \frac{\vec{n}\cdot \vec{\sigma}}{2}\right]_{A,k} \exp\left[i \theta \frac{\vec{n}\cdot\vec{\sigma}}{2}\right]_{B,k},
\end{equation}
we can interpret the model of Eq.~(\ref{eq:AKLT like}) as a non-trivial SPT phase protected by $SO(3)$ which has a $\hgcinput{SO(3)} \cong \ztwo$ classification. We now use the following extension to unwind the model:
\begin{equation} \label{eq: SO(3) extension}
1  \longrightarrow \ztwo \overset{i}{\longrightarrow} SU(2) \overset{s}{\longrightarrow} SO(3) \longrightarrow 1. 
\end{equation}
In order to make the system transform faithfully under $SU(2)$, we introduce an additional spin-$\frac{1}{2}$ particle at each site, which we will label $C$ as shown in Fig.~\ref{fig:AKLT_extended}. We extend $H$ with a trivial $SU(2)$ invariant Hamiltonian such that the ground state of the additional spins is a product of dimers of $SU(2)$ singlets: 
\begin{eqnarray}
\ket{\tilde{G}} &=& \ket{G} \otimes \prod_{\text{odd}~k} -\ket{\psi}_{Ck,Ck+1} = \ket{G} \otimes \prod_{\text{odd}~k} \left(\frac{\ket{\downarrow}_{C,k}\ket{\uparrow}_{C,k+1}-\ket{\uparrow}_{C,k}\ket{\downarrow}_{C,k+1}}{\sqrt{2}}\right), \\
\tilde{H} &=& H - \sum_{\text{odd~}k} \outerproduct{\psi}{\psi}_{Ck,Ck+1}. 
\end{eqnarray}

The on-site Hilbert space now transforms as the $\frac{1}{2} \otimes\frac{1}{2} \otimes\frac{1}{2} \cong \frac{3}{2} \oplus \frac{1}{2} \oplus \frac{1}{2}$ representation, which is faithful to $SU(2)$. It can be checked that the symmetry representation commutes with the extended Hamiltonian $\tilde{H}$ and leaves the ground state $\ket{\tilde{G}}$ invariant:
\begin{eqnarray}
\tilde{U}(\theta) &=& \prod_{k }  \exp\left[ i \theta \frac{\vec{n}\cdot\vec{\sigma}}{2}\right]_{A,k} \exp\left[i \theta \frac{\vec{n}\cdot\vec{\sigma}}{2}\right]_{B,k} \exp\left[ i \theta \frac{\vec{n}\cdot\vec{\sigma}}{2}\right]_{C,k}.
\end{eqnarray}

\begin{figure}[!htbp]
	\centering	
	\includegraphics[width=100mm]{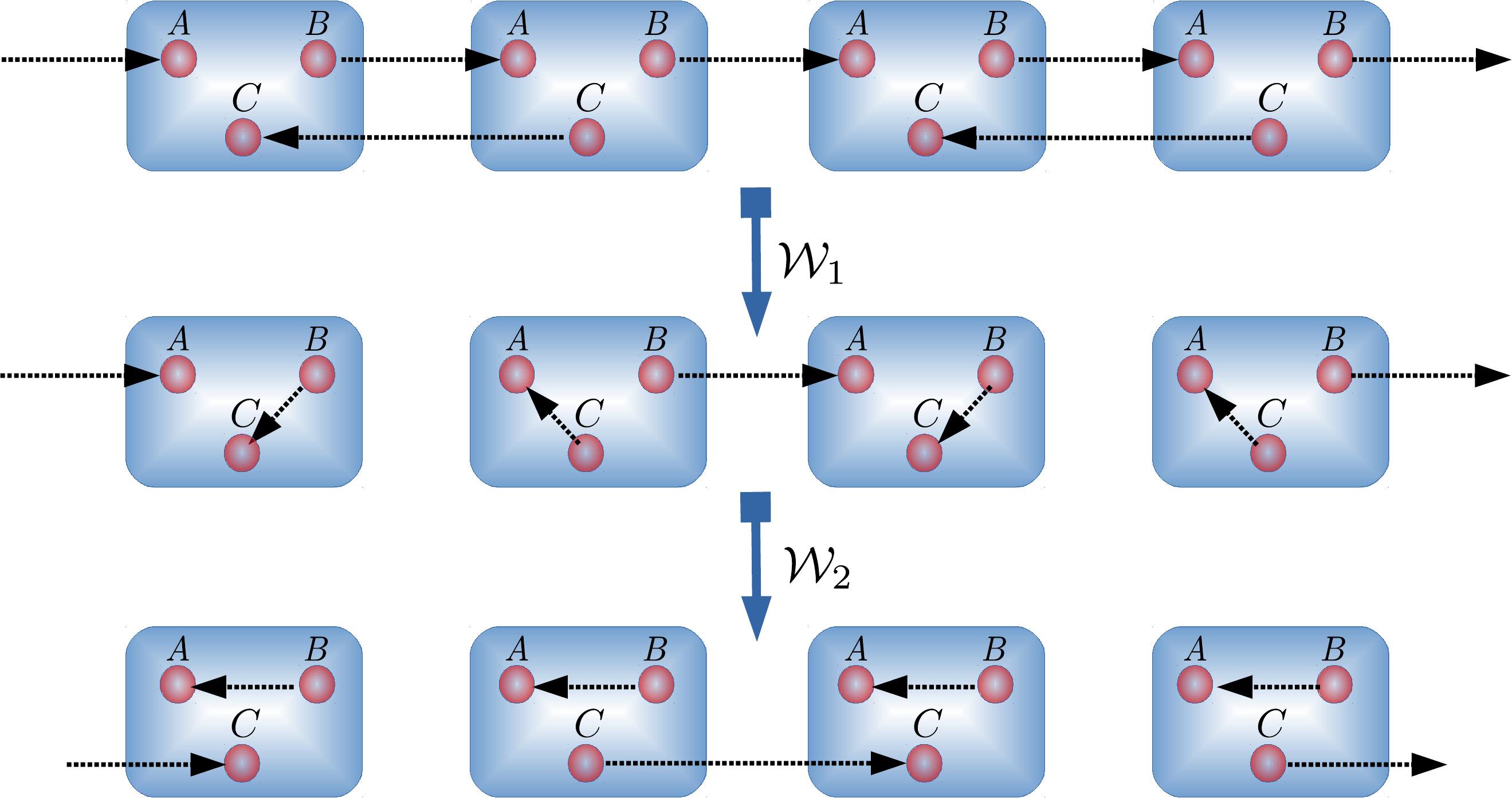}
	\caption{Unwinding of the AKLT-like model. \label{fig:AKLT_trivialize}}
\end{figure}

We use the two-layer FDUC $\mathcal{W} = \mathcal{W}_2 \mathcal{W}_1$ constructed using a series of entanglement swap operations to trivialize the system as shown in Fig.~\ref{fig:AKLT_trivialize}
\begin{eqnarray}
\mathcal{W}_1 &=& \prod_{\text{odd~} k} \mathcal{S}_{Ck,Ak+1}, \\
\mathcal{W}_2 &=& \prod_{\text{odd~} k} \mathcal{S}_{Ck,Ak} \prod_{\text{even~} k} \mathcal{S}_{Ck,Bk}, \\
\mathcal{S}_{AB} &=& \sum_{\alpha = \uparrow, \downarrow} \sum_{\beta = \uparrow, \downarrow}~ \outerproduct{\alpha}{\beta}_A \outerproduct{\beta}{\alpha}_B = \frac{1}{2} \left(\mathbb{1} + \vec{\sigma}_A. \vec{\sigma}_B\right).
\end{eqnarray}
Each two-qubit swap operator, $\mathcal{S}_{AB}$ is manifestly $SU(2)$ invariant and, as a result, so are $\mathcal{W}_1$ and $\mathcal{W}_2$. Altogether, $\mathcal{W}$ maps $\ket{\tilde{G}}$ and $\tilde{H}$ to the following trivial ground state, $\ket{G_0}$ and Hamiltonian $H_0$, thereby unwinding the SPT phase.
\begin{eqnarray}
\mathcal{W} \ket{\tilde{G}} &=& \prod_{k }  -\ket{\psi}_{Ak,Bk} \otimes \prod_{\text{even}~k} \ket{\psi}_{Ck,Ck+1} = \ket{G_0},\\
\mathcal{W} \tilde{H} \mathcal{W}^\dagger &=& -\sum_{k }  \outerproduct{\psi}{\psi}_{Ak,Bk} -\sum_{\text{even~} k} \outerproduct{\psi}{\psi}_{Ck,Ck+1}= H_0.
\end{eqnarray}

\subsubsection{As a time-reversal $ \mathbb{Z}_2^T$-invariant SPT phase}
Let us now take the same model but consider it as an SPT protected by the anti-unitary time reversal symmetry, $\ztwo^T$ generated by 
\begin{equation}
\mathcal{T} = \prod_{k }  \exp\left[ \frac{i\pi \sigma_y}{2}\right]_{A,k} \exp\left[ \frac{i\pi \sigma_y}{2}\right]_{B,k} \mathcal{K},
\end{equation}
where, $\mathcal{K}$ is the complex conjugation operation, and disregarding all other symmetries. Since each site contains two spin-1/2 particles (A and B), it is clear that the time-reversal operator squares to identity locally, i.e. $\mathcal{T}^2=\mathbb{1}$.

We now use the following extension to trivialize the model
\begin{eqnarray}\label{eq:Time reversal extension}
1 \longrightarrow \ztwo \overset{i}{\longrightarrow} \mathbb{Z}_4^T \overset{s}{\longrightarrow} \ztwo^T \longrightarrow 1. 
\end{eqnarray}
It turns out that we can repurpose the unwinding procedure involving $SO(3)$ to $SU(2)$
extension and also define a $\ztwo^T$ to  $\mathbb{Z}_4^T$ 
extension, with the $\mathbb{Z}_4^T$ being generated by
\begin{equation} 
\tilde{\mathcal{T}} = \prod_{k }  \exp\left[ \frac{i\pi \sigma_y}{2}\right]_{A,k} \exp\left[ \frac{i\pi \sigma_y}{2}\right]_{B,k} \exp\left[ \frac{i\pi \sigma_y}{2}\right]_{C,k} \mathcal{K}.
\end{equation} 

It can be checked that, because of the extra spin-1/2 particle on each site, $\tilde{\mathcal{T}}^2 = -1$ locally on each site, which means $\tilde{\cT}$ is an order-4 group element and generates the $\mathbb{Z}_4^T$ symmetry that we seek. It can easily be checked that $\tilde{H}$ and $\ket{\tilde{G}}$ are invariant under $\tilde{\mathcal{T}}$ as are $\mathcal{W}_1$ and $\mathcal{W}_2$, respectively. Thus, using the FDUC $\mathcal{W} = \mathcal{W}_2 \mathcal{W}_1$, we obtain the trivial Hamiltonian and ground state just as before.

To summarize, we have demonstrated how we can trivialize the AKLT-like model by symmetry extension. When viewed as an SPT phase protected by $SO(3)$, it can be trivialized using extension of Eq.~\ref{eq: SO(3) extension} and when viewed as an SPT phase protected by time-reversal symmetry, it can be trivialized using extension of Eq.~\ref{eq:Time reversal extension}. 

\begin{figure}[!htbp]
	\centering	
	\includegraphics[width=100mm]{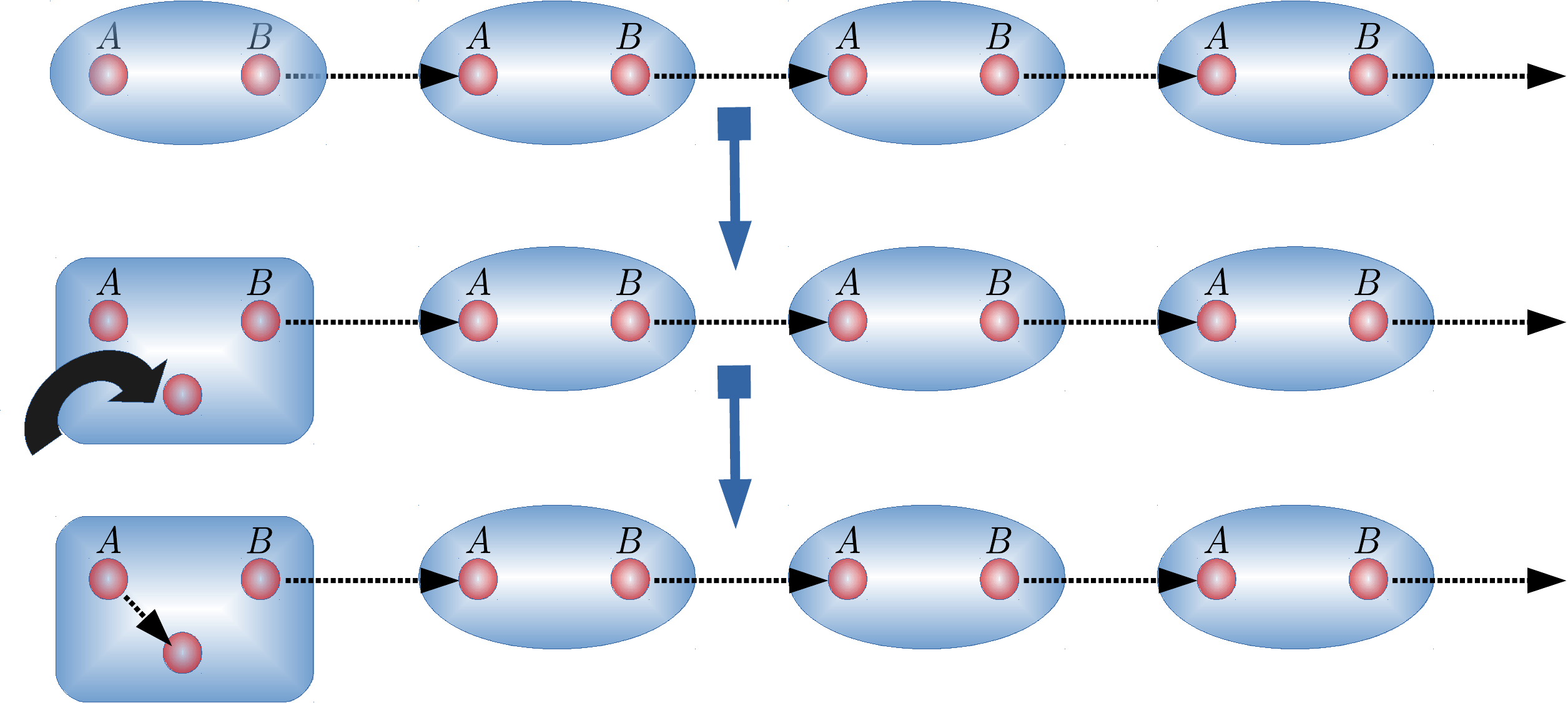}
	\caption{Gapping out the boundary modes by symmetry extension. \label{fig:AKLT_boundarygap}}
\end{figure}

For completeness, let us consider a simpler demonstration that this SPT phase can be trivialized by symmetry extension-- instead of unwinding the entire chain to a trivial one, we might be interested in simply gapping out the degenerate boundary spins by extending symmetry just on the boundary. This is very easy to do as shown in Ref.~\cite{WangWenWitten_PhysRevX.8.031048_symmetric}. Consider an open chain as shown in Fig.~\ref{fig:AKLT_boundarygap} with a dangling spin 1/2 at each end giving rise to a fourfold degeneracy. We can introduce additional spins that extends the symmetry on the boundary to $SU(2)$ and then tune in $SU(2)$ invariant boundary interaction terms, $h=-\outerproduct{\psi}{\psi}$ where $\ket{\psi}$ is the $SU(2)$ singlet, that favors entangling the two dangling spins into a singlet in the ground state thus lifting the degeneracy. This also applies to the interpretation of the boundary modes coming from time-reversal symmetry. Such a boundary gapping can be done for all the examples below but we will not mention it. We will focus on unwinding the entire system.   

\subsection{Unwinding the Cluster state}
\begin{figure}[!htbp]
	\centering	
	\includegraphics[width=100mm]{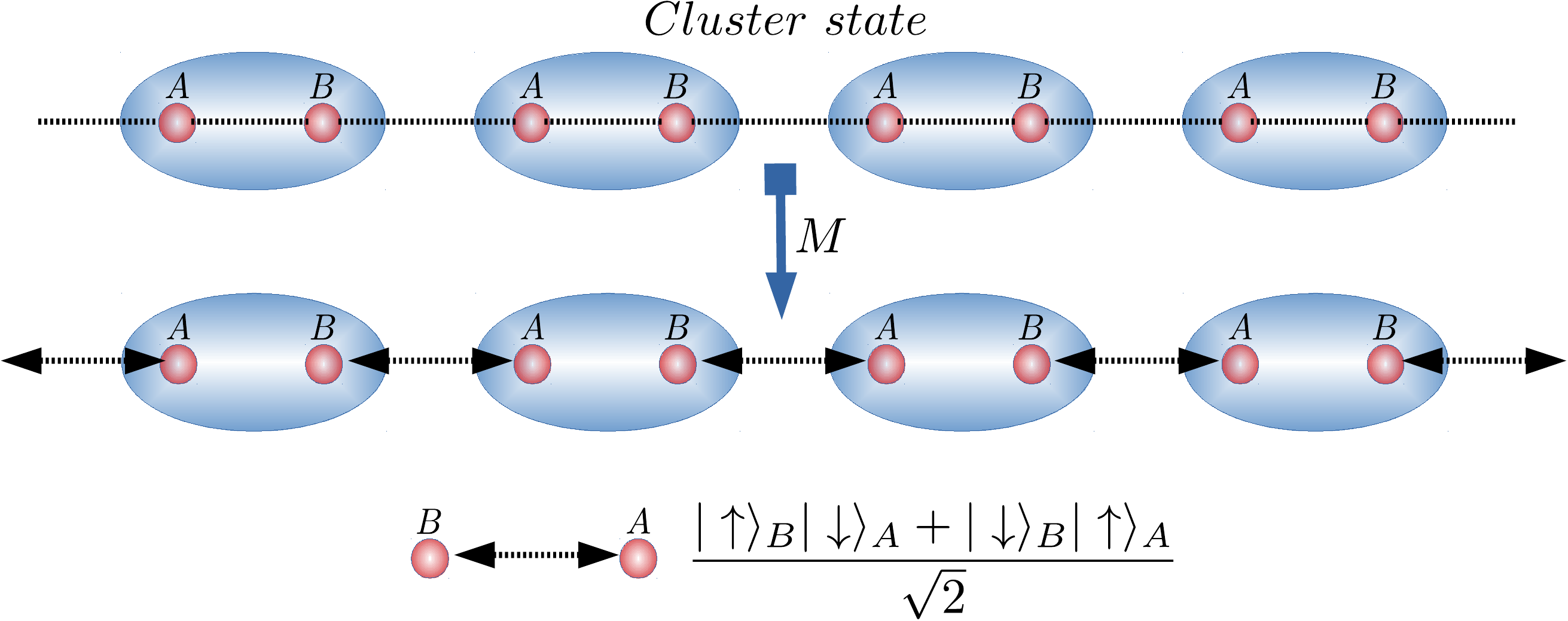}
	\caption{The cluster state before and after change of basis. \label{fig:Cluster}}
\end{figure}
We now consider another famous model of an SPT phase, the cluster state $\ket{\psi_C}$, and the Hamiltonian it is the ground state of, $H_c$:
\begin{eqnarray}
\ket{\psi_c} &=& \prod_{k }  CZ_{k,k+1}~ \prod_{j }  \ket{+}_j,\\
H_c &=& - \sum_{k }  \sigma^z_{k-1} \sigma^x_{k} \sigma^z_{k+1},
\end{eqnarray}
where, $\ket{+}$ is the positive eigenstate of $\sigma^x$ and $CZ_{ab}$ is the two-qubit operator 
\begin{equation}
CZ_{ab} = \frac{1}{2} \left(\mathbb{1} + \sigma^z_a + \sigma^z_b - \sigma^z_a \sigma^z_b \right)
\end{equation}
The cluster state~\cite{BreigelRaussendorf_2001PhysRevLett_Cluster} was introduced as a resource state for measurement-based quantum computation (MBQC)~\cite{RaussendorfBreigel_2001PhysRevLett_MBQC,RaussendorfBrownBreigel_2001PhysRevA_MBQC}. This model was later on understood to be a non-trivial SPT phase protected by a unitary on-site $\ztzt$ symmetry~\cite{Sonetal_2011EPL_ClusterSPT,SonAmicoVedral_2012QunatumInfoProcessing_ClusterSPT}, generated by the following two operators
\begin{eqnarray}
U(x) \equiv \prod_{\text{odd~} k}\sigma^x_k, ~U(z) \equiv \prod_{\text{even~} k}\sigma^x_k.
\end{eqnarray}
We also comment that the short-range entanglement structure that facilitates quantum computation is now understood as arising from the non-trivial SPT nature and the study of the utility of SPT phases for MBQC is a field of active research (see Refs.\cite{Miyake_2010PhysRevLett_SPTMBQC,ElseSchwarzBartlettDoherty_2012PhysRevLett_SPTMBQC,StephenEtal_2017PhysRevLett_SPTMBQC,Raussendorf_2017PhysRevA_SPTMBQC}).

For our purpose, it will be helpful to apply an on-site basis change to transform the cluster state into a more convenient form. First, let us collect two spins together and label them $A$ and $B$ to form a four dimensional local Hilbert space as shown in Fig.~\ref{fig:Cluster}. 
The symmetry generators can now be rewritten as
\begin{eqnarray}
U(x) = \prod_{k}\sigma^x_{A,k}, ~U(z) = \prod_{k}\sigma^x_{B,k}.
\end{eqnarray}

Next, we apply the on-site change of basis, $M$, defined as below to obtain the new form of the Hamiltonian, ground state, and symmetry generators: 
\begin{eqnarray}
M &=& \prod_{k }  \exp\left[ \frac{-i\pi \sigma^y}{4}\right]_{A,k} CZ_{Ak,Bk}, \\
M U(x) M^\dagger &\equiv& V(x) = \prod_{k }  \sigma^x_{A,k}~\sigma^x_{B,k},\\
M U(z) M^\dagger &\equiv& V(z) = \prod_{k }  i\sigma^z_{A,k}~ i\sigma^z_{B,k}, \\
M H_C M^\dagger &\equiv& H_{\mathcal{C}} = \sum_{k }  \left(\sigma^z_{B,i} \sigma^z_{A,i+1}- \sigma^x_{B,i} \sigma^x_{A,i+1}\right),\\
M \ket{\psi_C} M^\dagger &\equiv& \ket{\phi_{\mathcal{C}}} = \prod_{k }  \ket{\phi}_{Bk,Ak+1} =  \prod_{k }  \left(\frac{\ket{\uparrow}_{B,k}\ket{\downarrow}_{A,k+1}+ \ket{\downarrow}_{B,k}\ket{\uparrow}_{A,k+1}}{\sqrt{2}} \right).
\end{eqnarray}
 
 This is the same state that was briefly studied in Sec.~\ref{sec:three roads}. We now use the following symmetry extension to unwind this phase:
\begin{eqnarray}
1  \longrightarrow \ztwo \overset{i}{\longrightarrow} \mathbb{D}_8 \overset{s}{\longrightarrow} \ztzt \longrightarrow 1.
\end{eqnarray}
$\mathbb{D}_8$ is the order 8 dihedral group generated by two elements with the following presentation
\begin{equation} \label{eq:D8 presentation}
\mathbb{D}_8 = \innerproduct{a,x}{a^4 = x^2 = 1,x a x = a^{-1} }.
\end{equation} 

\begin{figure}[!htbp]
	\centering	
	\includegraphics[width=100mm]{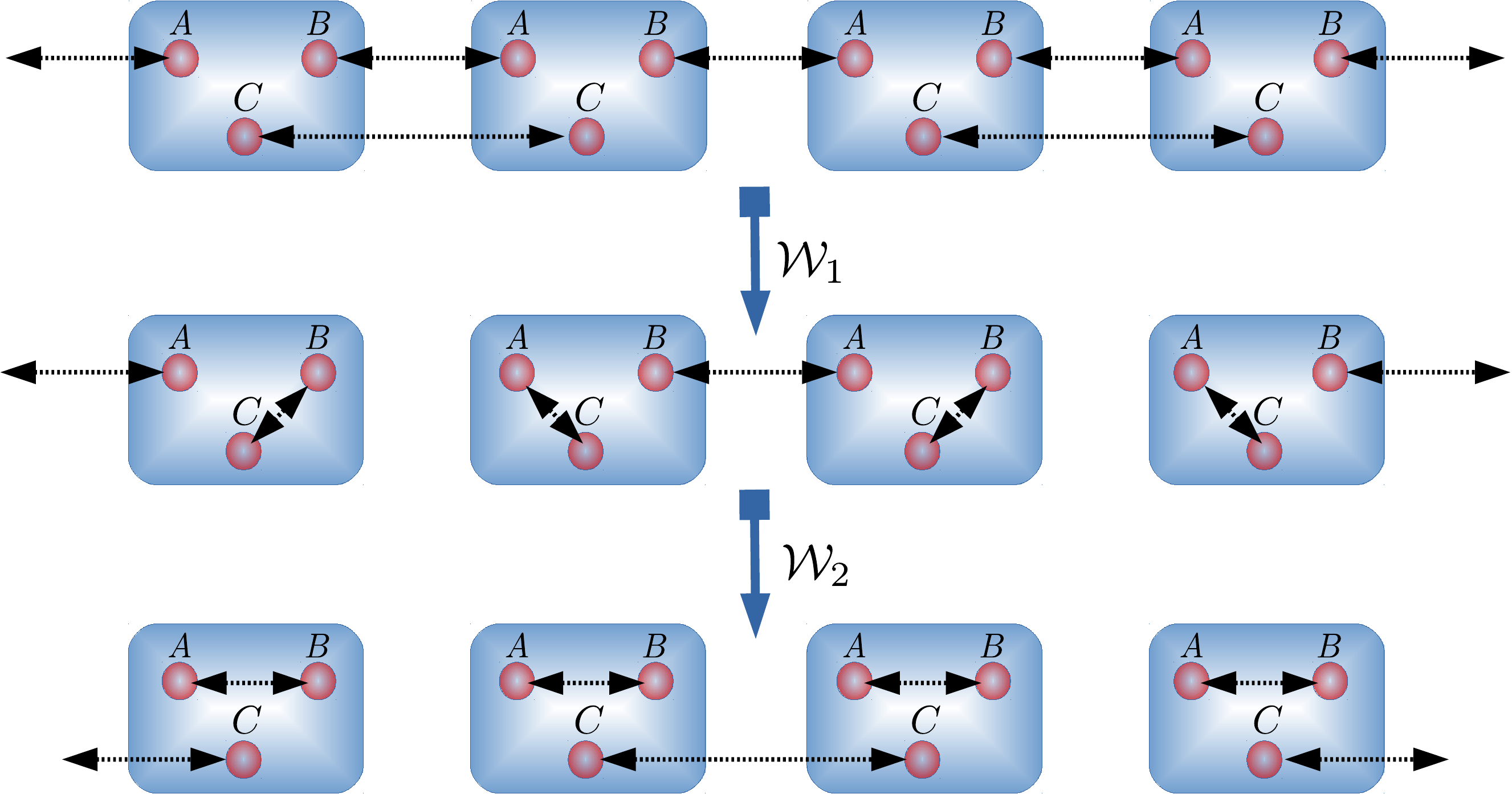}
	\caption{Unwinding the cluster state. \label{fig:Cluster_trivialize}}
\end{figure}

To achieve this, like before, we introduce a third qubit at each site, which we call $C$ and whose dynamics is governed by a dimerizing Hamiltonian that belongs to the trivial phase. The new ground state and Hamiltonian are as follows,
\begin{eqnarray}
\ket{\tilde{\phi}_{\mathcal{C}}} &=& \ket{\phi_\mathcal{C}} \otimes \prod_{\text{odd}~k} \ket{\phi}_{Ck,Ck+1} = \ket{G} \otimes \prod_{\text{odd}~k} \left(\frac{\ket{\downarrow}_{C,k}\ket{\uparrow}_{C,k+1}+\ket{\uparrow}_{C,k}\ket{\downarrow}_{C,k+1}}{\sqrt{2}}\right), \\
\tilde{H}_{\mathcal{C}} &=& H_{\mathcal{C}} + \sum_{\text{odd~} k} \left(\sigma^z_{C,i} \sigma^z_{C,i+1}- \sigma^x_{C,i} \sigma^x_{C,i+1}\right).
\end{eqnarray}
The symmetries of this model are generated by 
\begin{eqnarray}
\tilde{V}(x) &=& \prod_{k }  \sigma^x_{A,k}~\sigma^x_{B,k}~\sigma^x_{C,k},~~~\tilde{V}(z) = \prod_{k }  i\sigma^z_{A,k}~ i\sigma^z_{B,k}~i\sigma^z_{C,k}.
\end{eqnarray}
It can be checked that these generators satisfy the presentation of Eq.~\ref{eq:D8 presentation} and are a faithful representation of $\mathbb{D}_8$. With this, just like before, we can use a FDUC that commutes with this extended symmetry to unwind the system. In fact, we can use the exact same FDUC, $\mathcal{W} = \mathcal{W}_2 \mathcal{W}_1$ used in the previous section to do the job, as shown in fig.~\ref{fig:Cluster_trivialize}: 
\begin{eqnarray}
\mathcal{W}_1 &=& \prod_{\text{odd~} k} \mathcal{S}_{Ck,Ak+1}, \\
\mathcal{W}_2 &=& \prod_{\text{odd~} k} \mathcal{S}_{Ck,Ak} \prod_{\text{even~} k} \mathcal{S}_{Ck,Bk}, \\
\mathcal{S}_{AB} &=&\frac{1}{2} \left(\mathbb{1} + \vec{\sigma}_A. \vec{\sigma}_B\right).
\end{eqnarray}
Using this, we get the following trivial ground state and Hamiltonian
\begin{eqnarray}
\mathcal{W} \ket{\tilde{\phi}_\mathcal{C}} &=&  \ket{\phi_0} = \prod_{k }  \ket{\phi}_{Ak,Bk} \otimes \prod_{\text{even}~k} \ket{\phi}_{Ck,Ck+1}, \\
\mathcal{W} \tilde{H}_{\mathcal{C}} \mathcal{W}^\dagger &=&  H_0 =  \sum_{k }  \left(\sigma^z_{A,k} \sigma^z_{B,k}- \sigma^x_{A,k} \sigma^x_{B,k}\right) + \sum_{\text{even~} k} \left(\sigma^z_{C,k} \sigma^z_{C,k+1}- \sigma^x_{C,k} \sigma^x_{C,k+1}\right).
\end{eqnarray}

\subsection{General picture for finite on-site unitary symmetries: proof based on Schur cover}
\label{sec:Schur}

We now describe a general procedure to unwind fixed-point states of bosonic SPT phases in 1+1 D with any on-site unitary symmetry of a finite group, $G$ classified by $\omega \in \hgc$. First, in Sec.~\ref{sec:Schur_recipe}, we write down fixed-point SPT model Hamiltonians and ground states and provide an algorithm for extending the local Hilbert space and constructing the FDUC that unwinds these models. In Sec.~\ref{sec:Schur_connection}, we provide the connection to symmetry-extension and explain why the prescription of Sec.~\ref{sec:Schur_recipe} works.

\subsubsection{Algorithm to unwind fixed-point SPT states} \label{sec:Schur_recipe}
\begin{figure}[!htbp]
	\centering	
	\includegraphics[width=100mm]{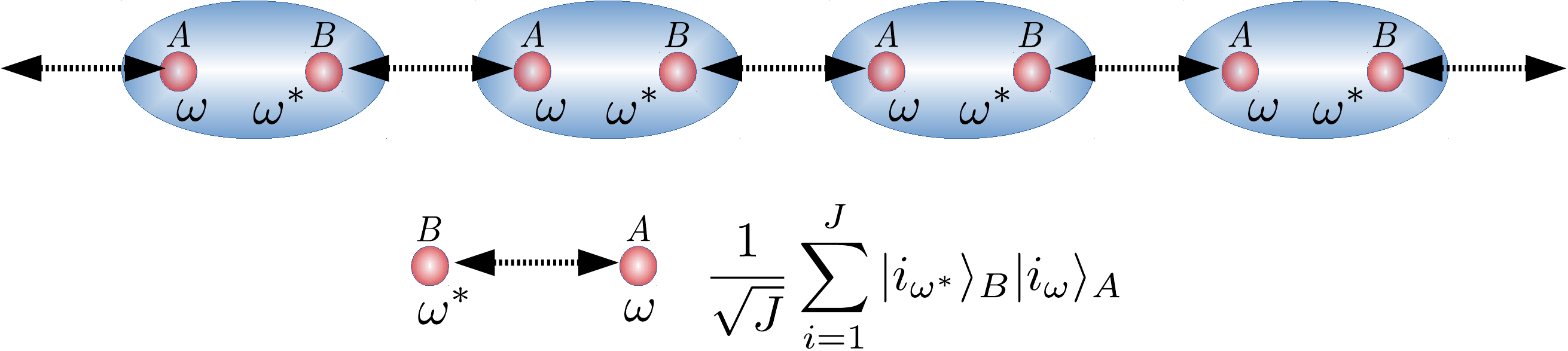}
	\caption{SPT state with finite on-site symmetry. \label{fig:General_onsite}}
\end{figure}

 We follow Ref.~\cite{ChenGuWen_2011PhysRevB_CompleteClassification1d} where it was shown that the classification of a bosonic SPT phase in 1+1 D corresponds to a classification of the projective representation of $G$ that the boundary degrees of freedom transform as. In particular, when $G$ is a finite on-site unitary symmetry, the classification of projective representations is in one-to-one correspondence with the elements of $\hgc$. Using this knowledge, we can write down fixed-point model for a 1+1 D bosonic SPT phases with an on-site unitary symmetry of a finite group $G$.  For the local on-site Hilbert space, we consider one spin that transforms as a projective representation belonging to class $\omega$ and another that transforms as $\omega^*$, the inverse of $\omega$ in the group $\hgc$. To be more precise, let $\ket{i_\omega} = \ket{1_\omega} \ldots \ket{J_\omega}$ be the basis states for some faithful $J$ dimensional projective representation of $G$ belonging to class $\omega \in \hgc$. Under group transformations, we have
\begin{eqnarray}
g:\ket{i_\omega} &\mapsto& \sum_{i' = 1}^{J} V(g)_{ii'} \ket{i'_\omega}, \\
V(g) V(h) &=& {\omega(g,h)} V(gh),
\end{eqnarray}
where $\omega(g,h)$ is a $U(1)$ phase factor. 
Now consider another spin of the same dimension $J$ that transforms as $\omega^*$, with basis states $\ket{i_{\omega^*}} = \ket{1_{\omega^*}} \ldots \ket{J_{\omega^*}}$ and the transformation property, 
\begin{eqnarray}
g:\ket{i_{\omega^*}} &\mapsto& \sum_{i' = 1}^{J} V^*(g)_{ii'} \ket{i'_{\omega^*}}, \\
V^*(g) V^*(h) &=&{{\omega^*}(g,h)} V^*(gh).
\end{eqnarray}
If we consider a physical site to contain both  spins, the representation of the symmetry that acts on the site, $U(g) \equiv V(g) \otimes V^*(g)$ can be checked to be a \emph{linear} representation of $G$ by observing that $U(g) U(h) = U(gh)$. To construct a non-trivial SPT state, we maximally entangle neighboring spins 
from different sites to form a symmetric state $\ket{\chi_{\omega}}$ as shown in Fig.~\ref{fig:General_onsite},
\begin{equation}
\ket{\chi_\omega}_{BA} = \frac{1}{\sqrt{J}} \sum_{i=1}^{J} \ket{i_{\omega^*}}_B \ket{i_\omega}_A.
\end{equation}

\begin{figure}[!htbp]
	\centering	
	\includegraphics[width=100mm]{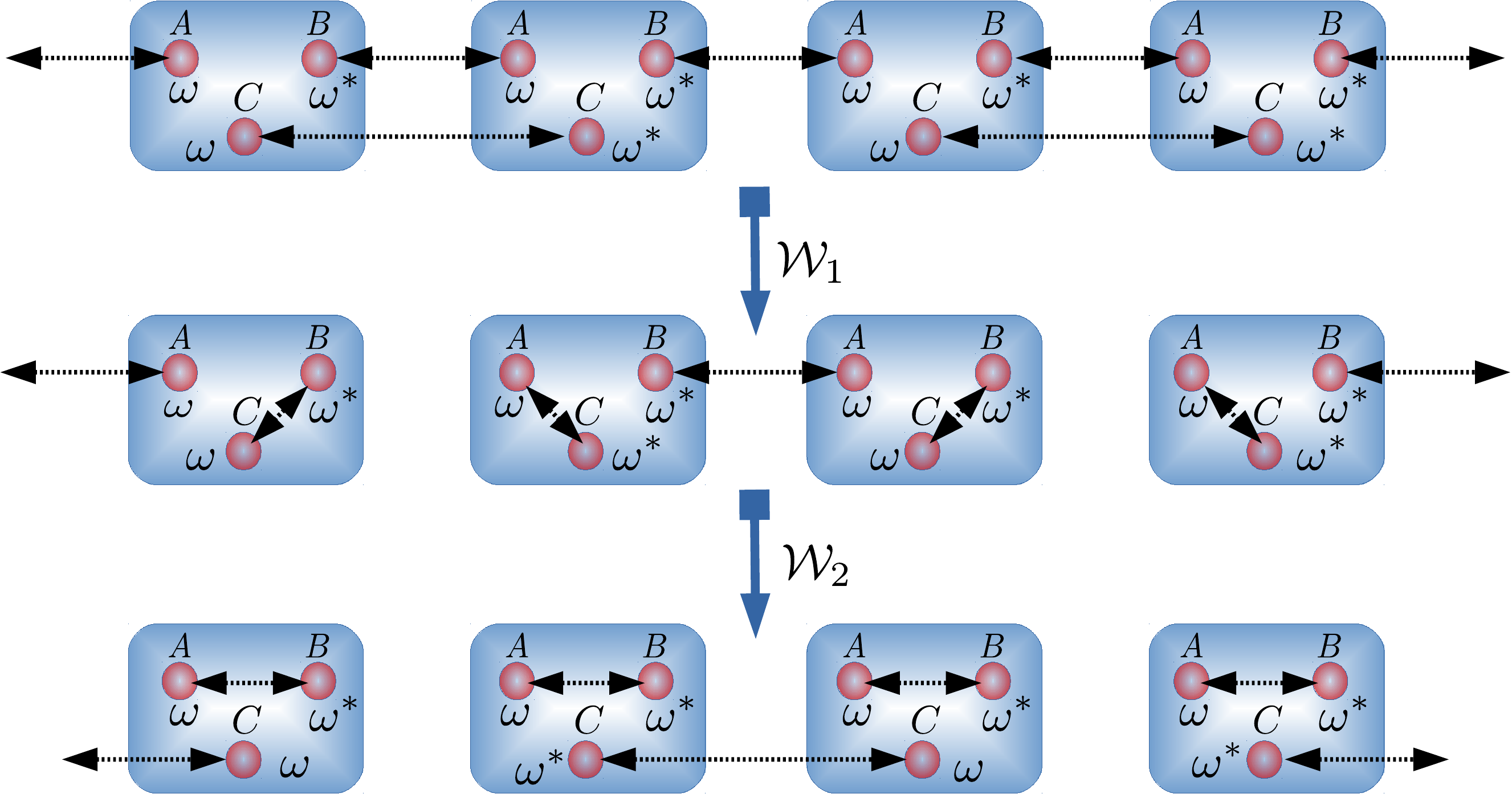}
	\caption{Unwinding SPT state with finite on-site symmetry. \label{fig:General_onsite_trivialize}}
\end{figure}

Using this, we can write down the following ground state and parent Hamiltonian:
\begin{eqnarray} \label{eq:fixed_pt_hamiltonian}
\ket{\psi_\omega} &=& \prod_k \ket{\chi_\omega}_{k k+1}, \\
H_\omega &=& - \sum_k \outerproduct{\chi_\omega}{\chi_\omega}_{kk+1}.
\end{eqnarray}
When defined on an open chain, it is easy to see that the model of Eq.~\ref{eq:fixed_pt_hamiltonian} has boundary degrees of freedom that transform as $\omega$ and $\omega^*$ projective representations of $G$ as expected.  We now demonstrate how to extend the local Hilbert space and trivialize the system. Consider an extension to the original system by introducing an ancillary degree of freedom, which we label $C$ and which transforms as $\omega$ and $\omega^*$ projective representations on alternating sites. With this extension, each site transforms as one of the following two projective representations of $G$:
\begin{eqnarray}
\tilde{U}_{\omega}(g) \equiv V(g) \otimes V^*(g) \otimes V(g) ~~~\text{or}~~~ \tilde{U}_{\omega^*}(g) \equiv V(g) \otimes V^*(g) \otimes V^*(g),
\end{eqnarray}
Let us also write down the ground state and Hamiltonian for the extended system
\begin{eqnarray}
\ket{\tilde{\psi}} &=& \ket{\psi_\omega} \prod_{\text{odd~}k} \ket{\chi_{\omega^*}}_{Ck Ck+1}, \\
\tilde{H} &=& H_\omega - \sum_{\text{odd~}k} \outerproduct{\chi_{\omega^*}}{\chi_{\omega^*}}_{CkCk+1}.
\end{eqnarray}
To trivialize the extended system, we use the following  swap operator
\begin{eqnarray}
\mathcal{S}^\omega_{AB} \equiv \sum_{i=1}^{J} \sum_{j=1}^{J} \outerproduct{i_\omega}{j_\omega}_A\otimes \outerproduct{j_\omega}{i_\omega}_B.
\end{eqnarray}
Finally, we define the following FDUC $\mathcal{W} = \mathcal{W}_2 \mathcal{W}_1$ to trivialize the system as shown in Fig.~\ref{fig:General_onsite_trivialize}, 
\begin{eqnarray}
\mathcal{W}_1 &=& \prod_{\text{odd~}k} \mathcal{S}^\omega_{C,k,A,k+1}, \\
\mathcal{W}_2 &=& \prod_{\text{odd~}k} \mathcal{S}^\omega_{A,k,C,k} \prod_{\text{even~}k} \mathcal{S}^{\omega^*}_{B,k,C,k}. 
\end{eqnarray}

\noindent Applying $\mathcal{W}$, we end up with the following trivial ground state and Hamiltonian
\begin{eqnarray}
\mathcal{W} \ket{\tilde{\psi}} &=& \prod_k \ket{\chi_{\omega^*}}_{AkBk}~ \prod_{\text{even~}k} \ket{\chi_{\omega}}_{Ck Ck+1},\\
\mathcal{W} \tilde{H} \mathcal{W}^\dagger &=& - \sum_k \outerproduct{\chi_{\omega^*}}{\chi_{\omega^*}}_{AkBk} - \sum_{\text{even~}k} \outerproduct{\chi_{\omega}}{\chi_{\omega}}_{CkCk+1}.
\end{eqnarray}

We have thus shown that for the fixed-point models of Eq.~\ref{eq:fixed_pt_hamiltonian}, we can extend the local Hilbert space of these models and unwind the SPT phase without making any reference to symmetry-extension! Below, in Sec.~\ref{sec:Schur_connection}, we explain how this is a special feature of 1+1 D SPT phases and explain the connection between projective representations and symmetry-extension.

\subsubsection{Connection to symmetry-extension}\label{sec:Schur_connection}
In Sec.~\ref{sec:Schur_recipe}, we found that fixed-point models of SPT phases defined on a local Hilbert space that transforms as a linear representation of a finite group $G$ can be unwound by extending the local Hilbert space in such a way that it transforms as a projective representation of $G$. It turns out that the extended local Hilbert space actually corresponds to the linear representation of a group $\tilde{G}$ which is an extension of $G$ of the kind we have been focusing on in this paper. We now discuss the relationship between the projective representations of $G$ and the extended symmetry group $\tilde{G}$, which unwinds the $G$ SPT phase. 

In general, for d+1 space-time dimensions, given $\mu \in \hdc$ that classifies an SPT phase, it is a difficult task to find the symmetry extension $\tilde{G}$ that will unwind the SPT phase. In 1+1 D, a given $\mu \in \hgc$ that classifies the SPT phase also classifies the projective representation of $G$ corresponding to the emergent boundary degrees of freedom. A trivial phase is one that has boundary degrees of freedom that transforms as linear representations of $G$. Consequently, if there exists a group $\tilde{G}$ such that the projective representation of $G$ corresponds to a linear one for $\tilde{G}$, such a group unwinds the $G$ SPT phase. The question is if there exists such a symmetry group. The answer is yes, as shown by Schur, and can be stated in the form of the following theorem~\cite{Schur_1904_SchuCover,Schur_1907_SchuCover}:

\emph{Theorem (Schur): Every finite group $G$ has associated to it at least one finite group  $\tilde{G}$, called a Schur cover, with the property that every projective representation of G can be lifted to an ordinary representation of $\tilde{G}$.}

This theorem is basically the specialization of the more general result of Ref.~\cite{WangWenWitten_PhysRevX.8.031048_symmetric} to 1+1 D. The interpretation in terms of projective representations is useful for our present purpose. The Schur cover, $\tilde{G}$ is precisely the extension that the extended local Hilbert space used in Sec.~\ref{sec:Schur_recipe} corresponds to. The advantage of the recipe of Sec.~\ref{sec:Schur_recipe} combined with the theorem by Schur is that we do not need to know the extended group $\tilde{G}$ to unwind the SPT phase by symmetry-extension. 

To understand this result better, let us look at the irreducible representations (irreps) of $\tilde{G}$. It can be shown~\cite{mendonca2017projective} that we can associate each such irrep $\Gamma_i$ with an element of the group $\mu \in \hgc$ as $\Gamma^\mu_i$. In particular, the irreps corresponding to the identity element of $\hgc$ contains the linear irreps of $G$ and in particular, the trivial irrep. The group structure of $\hdc$ is reflected in the Clebsch-Gordan decomposition of the direct product of irreps of $\tilde{G}$ as follows:
\begin{equation}
\Gamma^\mu_i \otimes \Gamma^\nu_j \cong \bigoplus_k N^{ij}_k \Gamma^{\mu\cdot\nu}_k,
\end{equation}
where, $N^{ij}_k$ is the multiplicity of irrep $\Gamma^{\mu\cdot\nu}_k$. In other words, the fusion of irreps corresponding to $\mu,~\nu \subset \hgc$ only produces irreps corresponding to $\mu\cdot\nu$. The representation with basis $\ket{i_\omega}$ is a direct sum of some $\tilde{G}$ irreps of class $[\omega]$. In the original system, the irrep content of the local Hilbert space corresponds to the class $[\omega]\cdot [\omega^*] = 1$ and hence is linear to the group $G$ as well as $\tilde{G}$ as seen above. However, the irrep content of the extended system correspond to the class $[\omega]\cdot[\omega^*]\cdot[\omega] = [\omega]$ or $[\omega]\cdot[\omega^*]\cdot[\omega^*] = [\omega^*]$ on alternating sites, which are both linear irreps only of $\tilde{G}$. 

This is a familiar story in the case of the irreps of $SU(2)$ which can be labeled by the total angular momentum quantum number $\frac{j}{2}$ where $j \in \mathbb{Z}$. We can divide these irreps into two classes depending on whether $j$ is even or odd corresponding to the elements of $\hgcinput{SO(3)} \cong\ztwo$ respectively. When $j$ is even, $\frac{j}{2}$ is an integer and is also an irrep of $SO(3)$. Thus, even/ odd $j$ irreps of $SU(2)$ are labeled by the trivial/ non-trivial element of $\hgcinput{SO(3)}$. Furthermore, the fusion outcome in the Clebsch-Gordan decomposition $\frac{j}{2} \otimes \frac{k}{2} \cong \frac{|j-k|}{2} \oplus \frac{|j-k|}{2} +1 \oplus \ldots \oplus \frac{|j+k|}{2}$ respects the $\ztwo$ structure of $\hgcinput{SO(3)}$ as defined above.

\section{Unwinding fermionic SPT phases: Class CII, AIII and BDI}
\label{sec:fermionic}
In this section, we present the unwinding of model Hamiltonians which realize certain short-range-entangled fermionic phases corresponding to three of the five Altland-Zirnbauer classes that have a non-trivial classification in the free limit in 1+1 D, namely, CII, AIII, and BDI. In particular, we focus on the fermion SPT phases which can be reinterpreted as bosonic ones where we can repeat the unwinding procedure of the previous section. We consider particular global symmetries of CII, AIII, and BDI, namely, $\frac{(\mathcal{U}(1) \rtimes \mathcal{Z}_4^C)}{\mathcal{Z}_2^f} \times \mathbb{Z}_2^T$, $\mathcal{U}(1) \times \ztwo^T$, $\ztwo^T \times \ztwof$ symmetries, respectively. We leave the question of unwinding inherently fermionic SPT phases to future work. 

A note about the notation used in describing global symmetries in fermionic systems-- any Hamiltonian describing the dynamics of fermions commutes with the fermion parity operator, $\pf = (-1)^{{N}_f}$. While this can be thought of as a symmetry, which we will call $\ztwof$, it is important to note that it can never be explicitly broken. One way to understand this is that this ``symmetry" is imposed by the condition of locality on the Hamiltonian. If we explicitly break $\ztwof$ by adding a term to the Hamiltonian that does not commute with $\pf$ like $\delta H = \sum_k \psi^{\dagger}_k + \psi_k$, the local terms in the Hamiltonian that are far-separated no longer commute, rendering the Hamiltonian non-local. Hence, the $\ztwof$ symmetry is sometimes implicitly assumed when defining global symmetries in the literature. In this paper however, we choose to list $\ztwof$ explicitly for clarity to avoid any potential confusion. Furthermore, whenever $\ztwof$ is part of a symmetry group, we indicate it using a ``mathcal" font. 
\subsection{Model Hamiltonians and their symmetries}
\begin{figure}[!htbp]
	\centering	
	\includegraphics[width=100mm]{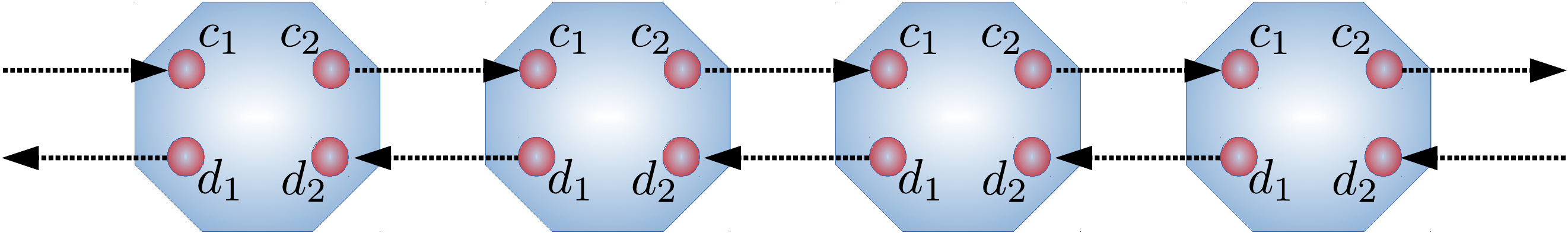}
	\caption{Generating Hamiltonian for fermion SPT phases in consideration. \label{fig:GeneratingHamiltonian}}
\end{figure}
The subset of fermionic SPT phases that we will be focusing on are all generated by the following Hamiltonian, shown in fig.~\ref{fig:GeneratingHamiltonian} consisting of two complex fermions per unit site. 
\begin{equation} 
H^2 = i \sum_{k } \left(c_{2,k} c_{1,k+1} - d_{2,k} d_{1,k+1}\right),\label{eq:GeneratingHamiltonian}
\end{equation} 
We represent a single complex fermion $i$ as a pair of Majorana fermions $c_i$, $d_i$ satisfying the relation
\begin{eqnarray}
\{c_i,c_j\} &=& \{d_i,d_j\} = 2 \delta_{ij} \\
\{c_i,d_j\} &=& 0
\end{eqnarray}
The Hamiltonian of Eq.~\ref{eq:GeneratingHamiltonian} was constructed by taking two layers of the so-called Kitaev chain~\cite{Kitaev_2001PhysicsUspekhi_KitaevChain} and performing a change of basis (see Appendices.~\ref{sec:AIII}, and \ref{sec:D}). By considering multiple layers of Hamiltonian of Eq.~\ref{eq:GeneratingHamiltonian}, as shown below, we can  get representatives of many phases. We direct the interested reader to Appendix.~\ref{app:fermion_hamiltonian_stracking} where we list representative model Hamiltonians for all non-trivial 1+1 D fermionic SPT phases which realize short-range-entangled fermionic phases corresponding to the five Altland-Zirnbauer classes that have a non-trivial classification in the free limit of which the Hamiltonians considered here are a subset. We consider $m$ copies of Hamiltonian.~\ref{eq:GeneratingHamiltonian}: 
\begin{equation} 
H^{2m} = i \sum_{\sigma=1}^m \sum_{k } \left(c_{\sigma,2,k} c_{\sigma,1,k+1} - d_{\sigma,2,k} d_{\sigma,1,k+1}\right),\label{eq:GeneratingHamiltonianLayers} 
\end{equation}  
\noindent which have the following symmetries that are important for our considerations:
\begin{enumerate}
	\item Fermion parity $\mathcal{Z}_2^f$ generated by $\pf = \prod_k \prod_{a=1}^2 \prod_{\sigma=1}^m (i c_{\sigma,a,k} d_{\sigma,a,k})$.
	\item Anti-unitary $\ztwo^T$  generated by $\mathcal{S} = \prod_k \prod_{\sigma=1}^m \left(c_{\sigma,2,k} d_{\sigma,1,k}\right) \mathcal{K}$ \text{, where} $\mathcal{K}$ denotes complex conjugation.
	\item Unitary $\mathcal{U}(1)$ with elements $V(\theta) = \prod_{k} \prod_{\sigma=1}^m  \exp{\frac{\theta}{2}\left(c_{\sigma,1,k} d_{\sigma,1,k} - c_{\sigma,2,k} d_{\sigma,2,k}  \right)}$.
	\item Unitary $\mathcal{Z}_4^C$ generated by $\mathcal{C} = \prod_{k } \prod_{\sigma=1}^m  \exp{\frac{\pi}{4} \sum_{a=1}^2 \left(c_{\sigma,a,k} c_{\sigma,a+1,k}-d_{\sigma,a,k} d_{\sigma,a+1,k}\right)}.$
\end{enumerate}
Let us list the SPT phases that the Hamiltonians of Eq.~\ref{eq:GeneratingHamiltonianLayers} represent.
\begin{itemize}
	\item $\{H^2,H^4,H^6\}$ belong to the non-trivial even numbered phases, $\nu = 2,4,6$ in the $\mathbb{Z}_8 = \{0,1,\ldots,7\}$ classification of class BDI with symmetry group $\ztwo^T \times \mathcal{Z}_2^f$.	
	\item $\{H^2,H^4,H^6\}$ belong to the non-trivial phases $\nu = 1,2,3$ in the $\mathbb{Z}_4 = \{0,1,2,3\}$ classification of class AIII with symmetry group $\mathcal{U}(1) \times \ztwo^T$.	
	\item $H^4$ belongs to the non-trivial phase in the $\ztwo= \{0,1\}$ classification of class CII with symmetry group $\frac{(\mathcal{U}(1) \rtimes \mathcal{Z}_4^C)}{\mathcal{Z}_2^f} \times \mathbb{Z}_2^T $.
\end{itemize}

\subsection{Unwinding m=2 model Hamiltonian}

\label{sec:3.2.unwind-4K}

\begin{figure}[!htbp]
	\centering	
	\includegraphics[width=110mm]{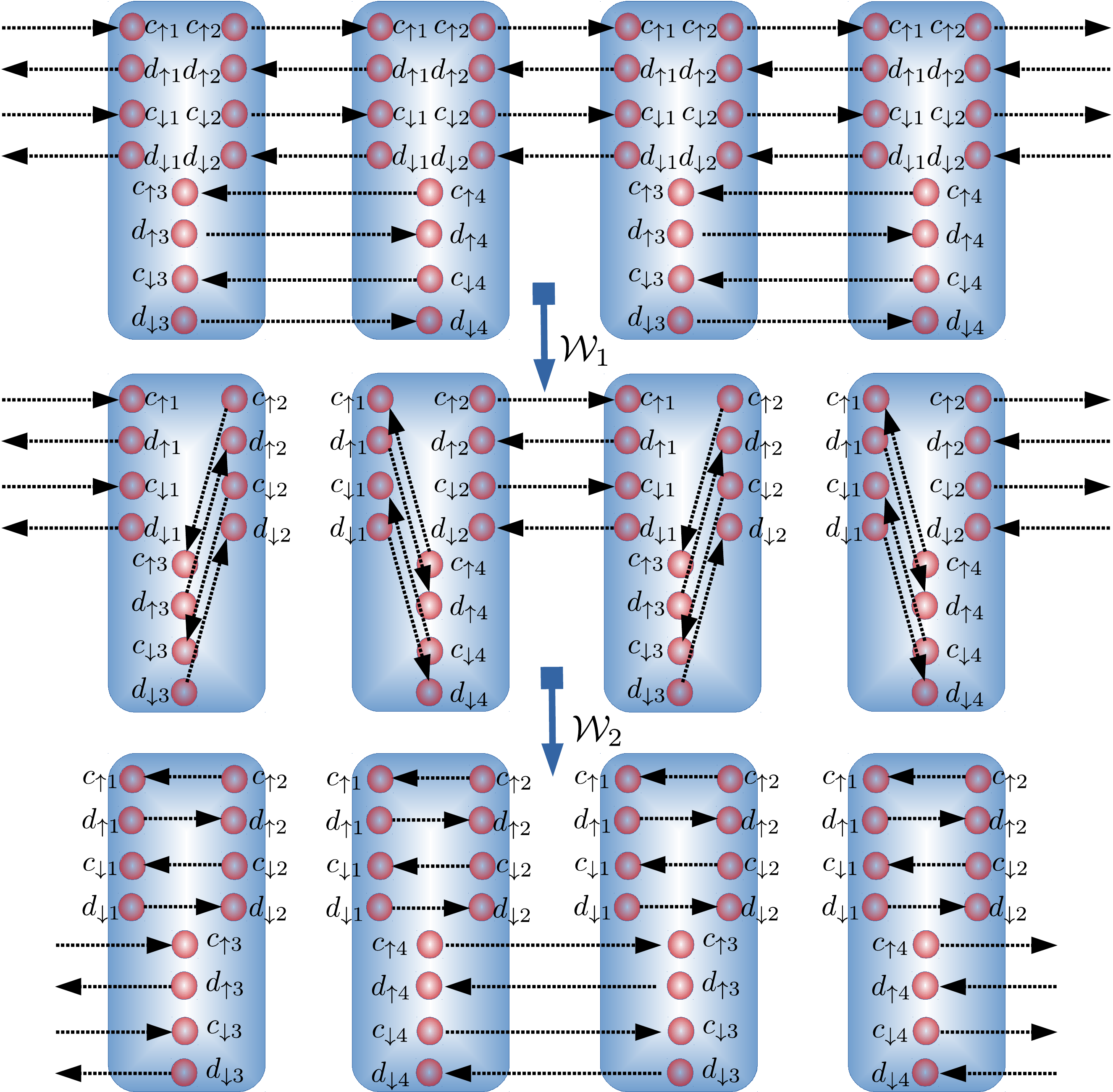}
	\caption{Trivialization of the non-trivial CII chain. \label{fig:CII_trivialize}}
\end{figure}

In this section we consider the fermion SPT phases that correspond to the $m=2$ Hamiltonian $H^4$ of Eq.~\ref{eq:GeneratingHamiltonianLayers}. We label the two layers as $\sigma = \uparrow, \downarrow$. Even though these are fermionic SPT phases, it has been understood that the non-trivial SPT nature for these phases can be understood as bosonic SPT phases belonging to Haldane phase~\cite{Fidkowski_Kitaev_2011PhysRevB_Classification1d,Verresen_Moessner_Pollmann_2017PRB_SPTtransition}. We trivialize this using an extension that was used before for the bosonic SPT phases-- that is, we extend the anti-unitary $\ztwo^T$  part of the symmetry to $\mathbb{Z}_4^T$ and leave the other symmetry generators unchanged. 
\begin{equation}
1  \longrightarrow \ztwo \overset{i}{\longrightarrow} \mathbb{Z}_4^T \overset{s}{\longrightarrow}  \mathbb{Z}_2^T \longrightarrow 1. 
\end{equation}
Note that the symmetry groups described in the previous section for various symmetry classes have the following embedding
\begin{eqnarray}
\text{CII}~\left(
{\frac{(\mathcal{U}(1) \rtimes \mathcal{Z}_4^C)}{\mathcal{Z}_2^f} \times \mathbb{Z}_2^T}\right) \xrightarrow[\mathcal{Z}_4^C]{\text{Disregard}} \text{AIII}~( \mathcal{U}(1)  \times \mathbb{Z}_2^T) \xrightarrow[\mathcal{U}(1)]{\text{Disregard}} \text{BDI}~(\ztwo^T \times \mathcal{Z}_2^f). \nonumber
\end{eqnarray}
As a result by disregarding successive symmetries as mentioned above, trivializing $H^4$ results in trivializing the only non-trivial SPT phase of Class CII, the $\nu=2$ SPT phase in the $\mathbb{Z}_4$ classification of class AIII and the $\nu=4$ SPT phase in the $\mathbb{Z}_8$ classification of class BDI. Let us now go into the details of how this is achieved. As we did for the bosonic case, we add additional degrees of freedom corresponding to two extra fermions per unit site. We will label the Majorana operators that correspond to these as $c_{3,\sigma,k}, d_{3,\sigma,k}$ for odd sites $k$ and $c_{4,\sigma,k}, d_{4,\sigma,k}$ for even sites $k$. We will see that this makes the local Hilbert space transform as a faithful representation of the extended symmetry  $\tilde{\mathcal{G}}= \left(
{\frac{(\mathcal{U}(1) \rtimes \mathcal{Z}_4^C)}{\mathcal{Z}_2^f} \times \mathbb{Z}_4^T}\right)$ for class CII (the extended symmetry for other classes can be obtained by disregarding symmetries as prescribed above). Furthermore, we add terms to the Hamiltonian $H^4$ corresponding to a trivial dimerized state for the new degrees of freedom. The new Hamiltonian and symmetry operators are
\begin{eqnarray}
\tilde{H}^4 &=& i \sum_{\sigma = \uparrow, \downarrow} \left(\sum_{k} \left(c_{\sigma,2,k} c_{\sigma,1,k+1} - d_{\sigma,2,k} d_{\sigma,1,k+1}\right) - \sum_{\text{odd}~k} \left(c_{\sigma,3,k} c_{\sigma,4,k+1} - d_{\sigma,3,k} d_{\sigma,4,k+1}\right) \right), \nonumber \\
\tilde{\mathcal{S}} &=& \prod_{\text{odd }k} \prod_{\sigma = \uparrow, \downarrow} i \left(c_{\sigma,2,k} d_{\sigma,1,k} d_{3,\sigma,k}\right)~~~\prod_{\text{even }k} \prod_{\sigma = \uparrow, \downarrow} i \left(c_{\sigma,2,k} d_{\sigma,1,k} c_{4,\sigma,k}\right)~~~\mathcal{K}, \nonumber  \\
V(\theta) &=&  \prod_{\text{odd } k} \exp{\frac{\theta}{2} \left( \sum_{\sigma=\uparrow,\downarrow} \sum_{a=1,2,3} (-1)^{a+1} c_{\sigma, a,k} d_{\sigma, a,k}\right)} ~~\prod_{\text{even } k} \exp{\frac{\theta}{2}  \left( \sum_{\sigma=\uparrow,\downarrow} \sum_{a=1,2,4}(-1)^{a+1} c_{\sigma, a,k} d_{\sigma, a,k}\right)} \nonumber,  \\
\mathcal{C} &=& \prod_{\text{odd } k} \exp{\frac{\pi}{4} \left(\sum_{a=1,2,3} c_{\downarrow, a,k} c_{\uparrow, a,k} - d_{\downarrow, a,k} d_{\uparrow, a,k}\right)}~~\prod_{\text{even } k} \exp{\frac{\pi}{4} \left( \sum_{a=1,2,4} c_{\downarrow, a,k} c_{\uparrow, a,k} - d_{\downarrow, a,k} d_{\uparrow, a,k}\right)} \nonumber.
\end{eqnarray}

It can be seen that $\tilde{S}^2$ is \emph{locally} -1 on both even and odd sites and hence is an extension of the original symmetry. This system can be trivialized using a two-layer FDUC $\mathcal{W} = \mathcal{W}_2 \mathcal{W}_1$ as shown in Fig.~\ref{fig:CII_trivialize} where
\begin{eqnarray}
\mathcal{W}_1 &=& \prod_{\text{odd } k}\exp{-\frac{\pi}{4}\left( \sum_{\sigma = \uparrow, \downarrow}  c_{\sigma,3,k} c_{\sigma,1,k+1} + d_{\sigma,3,k} d_{\sigma,1,k+1}\right)}, \nonumber \\
\mathcal{W}_2 &=& \prod_{\text{odd } k}\exp{-\frac{\pi}{4} \left( \sum_{\sigma = \uparrow, \downarrow}  c_{\sigma,3,k} c_{\sigma,1,k} + d_{\sigma,3,k} d_{\sigma,1,k} \right)}  \prod_{\text{even } k} \exp{\frac{\pi}{4} \left( \sum_{\sigma = \uparrow, \downarrow}  c_{\sigma,4,k } c_{\sigma,2,k } + d_{\sigma,4,k } d_{\sigma,2,k } \right)} .\nonumber
\end{eqnarray}

With a bit of straightforward algebra, it can be checked that $\mathcal{W}_1$ and $\mathcal{W}_2$ commute with the symmetry generators and the application of this FDUC does indeed leave us with a trivial Hamiltonian.
\begin{equation}
\mathcal{W} \tilde{H}^4 \mathcal{W}^{-1} = i \sum_{\sigma = \uparrow, \downarrow} \left(\sum_{k} \left(c_{\sigma,1,k} c_{\sigma,2,k} - d_{\sigma,1,k} d_{\sigma,2,k}\right) - \sum_{\text{even}~k} \left(c_{\sigma,4,k} c_{\sigma,3,k+1} - d_{\sigma,4,k} d_{\sigma,3,k+1}\right) \right).
\end{equation}

We conclude this section by summarizing the result of symmetry extension presented above on the classification of fermionic SPT phases in 1+1D in table.~\ref{table:table1}.
\begin{table}[!htb]
	\begin{center}
		\begin{tabular}{|c|c|c|c|}
			\hline
			Cartan class & Symmetry group $G$ & Extended symmetry group $\tilde{G}$ & Reduction in classification \\
			\hhline{====}
			BDI & $\ztwo^T \times \ztwof$ & $\mathbb{Z}_4^T \times \ztwof$ & $\mathbb{Z}_8 \rightarrow \mathbb{Z}_4$ \\
			\hline
			AIII & $\mathcal{U}(1) \times \ztwo^T$ & $\mathcal{U}(1) \times \mathbb{Z}_4^T$ & $\mathbb{Z}_4 \rightarrow \ztwo$ \\
			\hline
			CII & $\frac{(\mathcal{U}(1) \rtimes \mathcal{Z}_4^C)}{\mathcal{Z}_2^f} \times \mathbb{Z}_2^T$ & $\frac{(\mathcal{U}(1) \rtimes \mathcal{Z}_4^C)}{\mathcal{Z}_2^f} \times \mathbb{Z}_4^T$ & $\ztwo \rightarrow 1$ \\
			\hline
		\end{tabular}
	\end{center}
	\caption{Summary of fermionic SPT phases and the change classification by symmetry extension. \label{table:table1}}
\end{table}

\subsection{Comments on unwinding inherently fermionic SPT phases}

We might ask what happens if we try to unwind other fermionic SPT phases which cannot be reinterpreted as bosonic SPT phases. Let us make a few comments in this regard by focusing on class BDI, with symmetry $\ztwo^T \times \mathcal{Z}_2^f$ which has a $\mathbb{Z}_8$ classification. Since we have shown the unwinding of the $\nu = 4$ model, we will focus on $\nu = 1,2,3,4$.

The odd members $\nu = 1, 3, 5, 7$ have an odd number of boundary Majorana modes. In particular, the $\nu = 1$ model corresponds to the Kitaev chain with Hamiltonian~\cite{Kitaev_2001PhysicsUspekhi_KitaevChain}.
\begin{equation} ~\label{eq:KitaevMajorana}
H^1 = i \sum_k d_k c_{k+1}.
\end{equation}
It was shown in Kitaev's original paper that the Hamiltonian of Eq.~\ref{eq:KitaevMajorana} can be transformed to the trivial Hamiltonian, 
\begin{equation}
H^0 = i \sum_k c_k d_{k},
\end{equation}
\noindent by the algebra automorphism $c_i \mapsto d_i$ and $d_i \mapsto c_{i+1}$. Such a transformation cannot be generated by FDUC even if we allow the addition of trivial degrees of freedom and impose absolutely no symmetry constraint. Thus, the $\nu = 1$ member cannot be unwound by symmetry-breaking or symmetry-extension. It can be unwound by inversion and hence is said to be an invertible topologically ordered system. The other odd members are generated by stacking the Hamiltonian of Eq.~\ref{eq:KitaevMajorana} to those of Eq.~\ref{eq:GeneratingHamiltonianLayers} and cannot be unwound by symmetry-breaking or symmetry-extension for the same reason.

\begin{figure}[!htbp]
	\centering	
	\includegraphics[width=110mm]{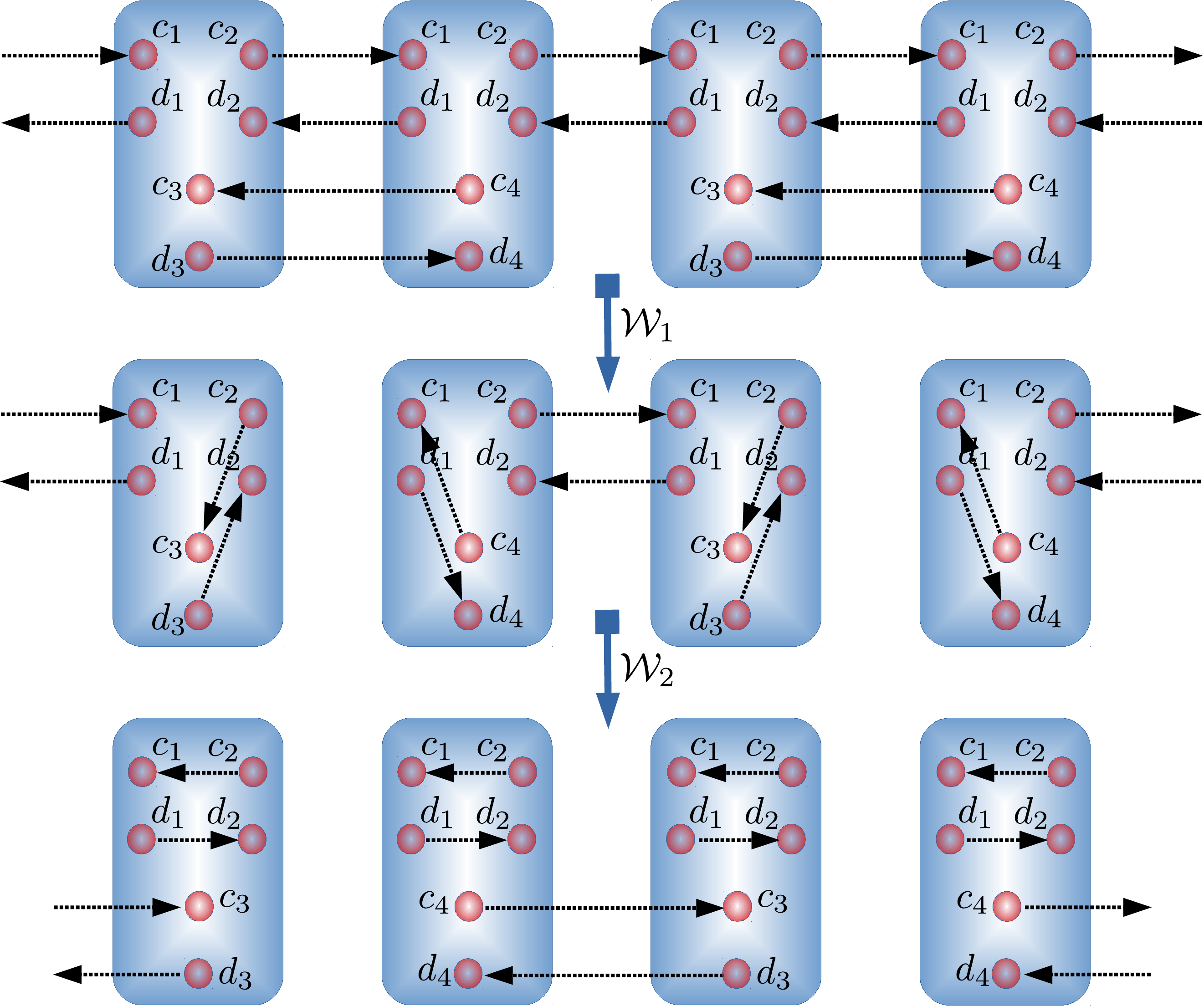}
	\caption{Trivialization of the $\nu=2$ BDI chain. \label{fig:BDI_trivialize}}
\end{figure}

We now focus on the even member $\nu = 2$ that the Hamiltonian of Eq.~\ref{eq:GeneratingHamiltonian} corresponds to. We could, in principle, use the same strategy as all other examples and unwind the model-- first by extending the local Hilbert space by adding one extra fermion, corresponding to operators $c_{3,k}, d_{3,k}$ on odd sites and $c_{4,k}, d_{4,k}$ on even sites as following:
\begin{equation} 
\tilde{H}^2 = i \sum_{k} \left(c_{2,k} c_{1,k+1} - d_{2,k} d_{1,k+1}\right)-i \sum_{\text{odd}~k} \left(c_{3,k} c_{4,k+1} - d_{3,k} d_{4,k+1}\right)
\end{equation} 
This system can be trivialized using a two-layer FDUC $\mathcal{W} = \mathcal{W}_2 \mathcal{W}_1$ as shown in Fig.~\ref{fig:BDI_trivialize} where
\begin{eqnarray}
\mathcal{W}_1 &=& \prod_{\text{odd } k}\exp{-\frac{\pi}{4}\left( c_{ 3,k} c_{ 1,k+1} + d_{ 3,k} d_{ 1,k+1}\right)}, \nonumber \\
\mathcal{W}_2 &=& \prod_{\text{odd } k}\exp{-\frac{\pi}{4} \left( c_{ 3,k} c_{ 1,k} + d_{ 3,k} d_{ 1,k} \right)}  \prod_{\text{even } k} \exp{\frac{\pi}{4} \left( c_{ 4,k } c_{ 2,k } + d_{ 4,k } d_{ 2,k } \right)} .\nonumber
\end{eqnarray}
However, the extended symmetry that leaves the FDUC invariant contains symmetry operators that do not commute with fermion parity! We leave a careful study of this unusual situation for future work.

\section{Conclusion and final remarks}
\label{sec:conclude}
\subsection{Comments on unwinding beyond fixed-point models}
In this paper we have demonstrated how fixed-point Hamiltonians belonging to non-trivial SPT phases can be unwound by symmetry-extension. We now comment on how this can be applied to Hamiltonians that are not fixed-point models. Recall that two Hamiltonians belong to the same SRE phase if their ground states can be mapped to each other using a FDUC, possibly with the addition of trivial degrees of freedom. Thus, given a Hamiltonian $H$ belonging to a certain SPT phase, we can unwind it, i.e., map its ground state to a product state in two steps:
\begin{enumerate}
	\item Use a FDUC, $\mathcal{F}_1$, possibly with the addition of trivial degrees of freedom or alternatively follow a path in Hamiltonian space adiabatically to map the ground state of $H$ to the ground state of a fixed-point Hamiltonian of the same SPT phase of the kind listed in Sec.~\ref{sec:bosonic} and Sec.~\ref{sec:fermionic}.
	\item Use a second FDUC, $\mathcal{F}_2$ of the kind constructed in Sec.~\ref{sec:bosonic} and Sec.~\ref{sec:fermionic} to unwind the fixed-point ground state to a trivial state. 
	\item Altogether $\mathcal{F}_2\mathcal{F}_1$ unwinds $H$ by symmetry-extension.
\end{enumerate}

\subsection{Symmetry-embedding, symmetry-extension and classifications}

\label{sec:6-1-web}

We can organize our result on 1+1D fermionic SPT states in terms of  the symmetry embedding web recently introduced in \cite{1711.11587.GPW}.
Table~\ref{table:web} shows the ten particular global symmetries (in terms of the Cartan notations)
and their symmetry embedding pattern.
The $G_1 \to G_2$ with an arrow connecting between groups means that
the symmetry group $G_1$ embeds $G_2$, or equivalently, the $G_1$ can be broken down to a subgroup $G_2$.
We focus on the five particular symmetry groups, 
$\ztwof$,
$\ztwo^T \times \ztwof$,
$\mathcal{Z}_4^T$, 
$\mathcal{U}(1) \times \ztwo^T$=$\frac{\CU(1)^{}_{} \times \cZ^{{T}}_{4}}{\cZ_2^f}$,
$\frac{(\mathcal{U}(1) \rtimes \mathcal{Z}_4^C)}{\mathcal{Z}_2^f} \times \mathbb{Z}_2^T$-symmetries
(or Cartan notations as D, BDI, DIII, AIII and CII),
that are marked with frame boxes in Table.~\ref{table:web}.

In Appendix.~\ref{app:fermion_hamiltonian_stracking}, we construct the above five particular symmetry groups of fermionic SPT states 
by stacking fermionic Kitaev chains. It is helpful to use topological invariants (i.e. SPT invariants)  to describe the ground states of these SPT states. Reference.~\cite{KapustinThorngrenWang_2015JHEP_Cobordism} points out that
D class with $\ztwof$-symmetry is characterized by the Arf invariant, 
BDI class with $\ztwo^T \times \ztwof$-symmetry
is characterized by the Arf-Brown-Kervaire invariant. 
Combining together with the information of stacking Kitaev chain constructions and
the topological invariants of these SPT states, we can summarize our finding as follows.

\begin{table*}[!h] 
	\centering 
	{
		\begin{center}
			\begin{tikzpicture}\kern-13mm[>=stealth,->,shorten >=2pt,looseness=.5,auto]
			\matrix (M)[matrix of math nodes,row sep=8mm,column sep=5mm]{
				& 
				\begin{minipage}[c]{.85in} C: \ccblue{$\mathcal{SU}(2)$},\\ 
				No class 
				\end{minipage} &    &  
				\begin{minipage}[c]{.58in} A: \ccblue{$\CU(1)$},\\ 
				No class  
				\end{minipage} 
				&    &  
				\fbox{\parbox{1.4cm}{
						\begin{minipage}[c]{.8in} D: \ccblue{$\CZ_2^f$}, \\ 
						$(\nu_{\rm D})$ 
						$\in$  \\
						$\Z_2$-class 
						\end{minipage} 
				}}
				\\
				\begin{minipage}[c]{1.in}
				CI: \\
				\ccblue{$\frac{\mathcal{SU}(2) \times \cZ_4^T}{\cZ_2^f}$}, \\
				$(\alpha)$ 
				$\in$ 
				$\Z_2$-class
				\end{minipage}  
				&   &  
				\begin{minipage}[c]{1.in}
				AI:\\
				\ccblue{$\CU(1)   \rtimes \Z^{{T}}_{2}$},\\
				$(\alpha)\in$
				$\Z_2$-class
				\end{minipage}
				&   &  
				\fbox{\parbox{1.8cm}{
						\begin{minipage}[c]{.8in}
						BDI:\\
						\ccblue{$\Z_2^T \times \cZ_2^f$},\\
						$(\nu_{\rm BDI})$ 
						$\in$  \\
						$\Z_8$-class
						\end{minipage}
				}}
				&\\
				&   & 
				\fbox{\parbox{3.cm}{
						{\begin{minipage}[c]{3.cm}
							AIII:\\
							\ccblue{$\frac{\CU(1)^{}_{} \times \cZ^{{T}}_{4}}{\cZ_2^f}$ $={\CU(1)^{}_{} \times \Z^{{T}}_{2}}$},\\
							$(\nu_{\text{AIII}}) \in$
							$\Z_4$-class
							\end{minipage} }
				}}
				&   & 
				\fbox{\parbox{2.9cm}{
						\begin{minipage}[c]{1.2in}
						\rm{DIII}:
						\ccblue{$\cZ_4^T $},\\
						$(\nu_{\rm DIII})\in $ 
						$\Z_{2}$-class  
						\end{minipage}
				}}
				\\
				\fbox{\parbox{3.cm}{
						\begin{minipage}[c]{1.1in}
						CII:\\
						\ccblue{$\frac{(\mathcal{U}(1) \rtimes \mathcal{Z}_4^C)}{\mathcal{Z}_2^f} \times \mathbb{Z}_2^T$},\\
						$(\nu_{\text{CII}})$ 
						$\in$  
						$\Z_2$-class
						\end{minipage}  
				}}
				&   & 
				\begin{minipage}[c]{1.in}
				AII: \ccblue{$\frac{\CU(1)   \rtimes \cZ^{{T}}_{4}}{\cZ_2^f}$},\\
				No class
				\end{minipage}  
				\\
			};
			\foreach \a/\b in {1-2/1-4, 1-4/1-6, 2-1/2-3, 2-3/2-5, 
				3-3/3-5, 4-1/4-3, 2-1/1-2, 2-3/1-4, 
				2-5/1-6, 3-3/2-5, 3-5/1-6, 4-1/1-2, 4-1/3-3, 
				4-3/3-5, 2-1/3-3, 3-3/1-4, 4-3/1-4} {
				\draw[thick,->](M-\a)--(M-\b);
			}
			\end{tikzpicture}
		\end{center}
	}
	\caption{
		The symmetry embedding web of 1+1D fermionic SPT states relevant for Cartan symmetry classes (See also \cite{1711.11587.GPW}, in particular for 3+1D cases).
		The web suggests the maps between the nontrivial classes of their classifications of SPT states (or topological terms).
		The web can also suggest a possible symmetry group extension to unwind the SPT states.
		For the 5 Cartan classes of SPT states in the boxed frames, we provide their lattice realizations in Appendix.~\ref{app:fermion_hamiltonian_stracking}.
		See the main text in Sec.~\ref{sec:6-1-web} for further discussions. 	\label{table:web}}
\end{table*}

\begin{enumerate}
	\item The D class ($\ztwof$-symmetry) is related to the Arf invariant and has a $ \Z_2$ classification. 
	We do \emph{not} find any \emph{symmetry extension} to trivialize this SPT state. 
	We cannot break $\ztwof$ symmetry thus we cannot unwind this SPT state by \emph{symmetry breaking}, either. 
	This is due to the fact that it has a robust \emph{invertible fermionic topological order} protected by no global symmetries except the $\ztwof$-symmetry.\footnote{
	In terms of Wen's definition, this single layer Kitaev chain is a long-range entangled state. See Sec.~\ref{sec:unwind-SRE-LRE} for more discussions.}
	For an open Kitaev chain, there are two dangling Majorana modes on the two edges. Thus, the ground state degeneracy (GSD) is 2.
	However, it is known that by stacking two such chains and adding interactions, we can obtain a trivial class with a trivial vacuum ground state and single ground state degeneracy, GSD=1.
	
	\item The BDI class ($\ztwo^T \times \ztwof$-symmetry) is related to the Arf-Brown-Kervaire invariant and has a $\Z_8$-classification. A single Kitaev chain represents $\nu_{\rm BDI} =1  \in \Z_8$-class and
	we can stack eight chains with interactions to get a trivial class without breaking the $\ztwo^T \times \ztwof$-symmetry.
	
	By \emph{symmetry breaking}, we can reduce BDI to D class,
	thus $\nu_{\rm D}  \in \Z_2$-class; namely, the even classes of $\nu_{\rm BDI} = \text{even}  \in \Z_8$-class
	become trivial once we break $\ztwo^T \times \ztwof$ to $\ztwof$-symmetry.
	
	By \emph{symmetry extension}, we can also trivialize the class corresponding to $\nu_{\rm BDI} = 4$. For a four-layer Kitaev chain, we can ``double'' the system to make it an eight-layer Kitaev chain, 
	which suggests that a four-layer Kitaev chain is a $\Z_2$ subclass. Thus, intuitively, a symmetry extension by $\Z_2$ may unwind the $\nu_{\rm BDI} =4  \in \Z_8$-class SPT state.
	Indeed, this observation agrees with Sec.~\ref{sec:3.2.unwind-4K}.

	\item The DIII class ($\cZ_4^T$-symmetry)
	has a
	$\nu_{\rm DIII}  \in \Z_2$-classification.
	A two-layer Kitaev chain can represent $\nu_{\rm DIII}=1  \in \Z_2$-class and
	we can stack 4 chains with interactions to get a trivial class without breaking the $\cZ_4^T$-symmetry.
	
	By \emph{symmetry breaking}, we can reduce DIII to D class,
	thus to $\nu_{\rm D}=0  \in \Z_2$-class; namely, \emph{all} classes of $\nu_{\rm DIII}   \in \Z_2$-class
	become trivial once we break $\cZ_4^T$ to $\cZ_2^f$-symmetry.

	\item The AIII class ({$\frac{\CU(1)^{}_{} \times \cZ^{{T}}_{4}}{\cZ_2^f}$ or its equivalent rewriting ${\CU(1)^{}_{} \times \Z^{{T}}_{2}}$}-symmetry )
	has a
	$\nu_{\rm AIII}  \in \Z_4$-classification.
	A two-layer Kitaev chain can represent $\nu_{\rm DIII} =1  \in \Z_4$-class,
	we can stack 8 chains with interactions to get a trivial class without breaking of its symmetry.
	
	By \emph{symmetry breaking}, we can reduce AIII to BDI or DIII class, then reduce further to D class.
	The classes in their classifications can be mapped easily. 
	
	By \emph{symmetry extension}, we can also trivialize the $\nu_{\rm AIII} = 2$ class. For a four-layer Kitaev chain, we can fold twice  the system to make it an 8-layer Kitaev chain, 
	which suggests that a four-layer Kitaev chain is a $\Z_2$ subclass. 
	Thus, intuitively, a symmetry extension by $\Z_2$ may unwind the $\nu_{\rm AIII} =2  \in \Z_4$-class SPT state.
	Indeed, in Sec.~\ref{sec:3.2.unwind-4K}, we find such a $\Z_2$-extension.

	\item The CII class $\left({\frac{(\mathcal{U}(1) \rtimes \mathcal{Z}_4^C)}{\mathcal{Z}_2^f} \times \mathbb{Z}_2^T} \right)$
	or its equivalent rewriting 
	{$\left(\frac{(\CU(1)   \times \Z^{{CT}}_{2} )\rtimes \cZ_4^C}{\cZ_2^f}\right)$}
	has a
	$\nu_{\rm CII}  \in \Z_2$ classification.
	A four-layer Kitaev chain can represent $\nu_{\rm CII} =1  \in \Z_2$-class,
	we can stack 8 chains with interactions to get a trivial class without breaking the any of its symmetry.
	
	By \emph{symmetry breaking}, we can reduce CII to AIII, thus to BDI or DIII class, then reduce further to D class.
	The classes in their classifications can be mapped easily. 
	
	By \emph{symmetry extension}, we can also trivialize all classes of $\nu_{\rm CII}   \in \Z_2$-class.
	For a four-layer Kitaev chain, we can ``double''  the system to make it an 8-layer Kitaev chain, 
	which suggests that a four-layer Kitaev chain is a $\Z_2$ subclass. 
	Thus, intuitively, a symmetry extension by $\Z_2$ may unwind the $\nu_{\rm CII} =1  \in \Z_2$-class SPT state.
	Indeed in Sec.~\ref{sec:3.2.unwind-4K}, we succeed to find such a $\Z_2$-extension.

\end{enumerate}

{
	We conclude this section by summarizing in tables.~(\ref{table:table3}-\ref{table:table5}), the result of symmetry extension presented above, on the classification of fermionic SPT phases in 1+1D
	presented in Secs.~\ref{sec:fermionic},
	and also the symmetry breaking and symmetry extension of fermionic SPT phases in 1+1D presented in Sec.~\ref{sec:three roads} and \ref{sec:bosonic}.
	\begin{table}[!htb]
		\begin{center}
			\begin{tabular}{|c|c|c|c|}
				\hline
				Cartan class & Symmetry group $G$ & Extended symmetry group $\tilde{G}$ & Reduced classification \\
				\hhline{====}
				BDI & $\ztwo^T \times \ztwof$ & $\mathbb{Z}_4^T \times \ztwof$ & $\mathbb{Z}_8 \rightarrow \mathbb{Z}_4$ \\
				\hline
				AIII & $\mathcal{U}(1) \times \ztwo^T$ & $\mathcal{U}(1) \times \mathbb{Z}_4^T$ & $\mathbb{Z}_4 \rightarrow \ztwo$ \\
				\hline
				CII & $\frac{(\mathcal{U}(1) \rtimes \mathcal{Z}_4^C)}{\mathcal{Z}_2^f} \times \mathbb{Z}_2^T$ & $\frac{(\mathcal{U}(1) \rtimes \mathcal{Z}_4^C)}{\mathcal{Z}_2^f} \times \mathbb{Z}_4^T$ & $\ztwo \rightarrow 1$ \\
				\hline
			\end{tabular}
		\end{center}
		\caption{Summary of unwinding 1+1D fermionic SPT phases and the change of classification by symmetry extension. 		\label{table:table3}}
	\end{table}
	\begin{table}[!htb]
		\begin{center}
			\begin{tabular}{|c|c|c|c|}
				\hline
				\multicolumn{4}{c}{Symmetry breaking $G$ to $G'$}  \\
				\hline\hline
				SPT phase & Symmetry group $G$ & Unbroken  subgroup ${G}'$ & Reduced classification \\
				\hhline{====}
				Haldane/AKLT chain & $\ztwo^T$ & $0$ & $\mathbb{Z}_2 \rightarrow 1$ \\
				\hline
				Haldane/AKLT chain  & $SO(3)$ or $\Z_2\times \Z_2$  & $0$ or $\Z_2$  & $\mathbb{Z}_2 \rightarrow 1$ \\
				\hline
				Cluster state  & $\Z_2\times \Z_2$  & $0$ or  $\ztwo$  & $\mathbb{Z}_2 \rightarrow 1$		\\
				\hline
			\end{tabular}
		\end{center}
		\caption{Summary of unwinding 1+1D bosonic SPT phases by symmetry breaking studied in Sec.~\ref{sec:three roads} 		\label{table:table4}}
	\end{table}
	\begin{table}[!htb]
		\begin{center}
			\begin{tabular}{|c|c|c|c|}
				\hline
				\multicolumn{4}{c}{Symmetry extension $1  \longrightarrow {K}  \overset{i}{\longrightarrow} \tilde{G} \overset{s}{\longrightarrow}  G \longrightarrow 1$}  \\
				\hline\hline
				SPT phase & Symmetry group $G$ & Extended symmetry $\tilde{G}$ & Reduced classification \\
				\hhline{====}
				Haldane/AKLT chain & $\ztwo^T$ & $\mathbb{Z}_4^T$ & $\mathbb{Z}_2 \rightarrow 1$ \\
				\hline
				Haldane/AKLT chain  & $SO(3)$  & $SU(2)$   & $\mathbb{Z}_2 \rightarrow 1$ \\
				\hline
				Cluster state  & $\Z_2\times \Z_2$  &  $\mathbb{D}_8$  & $\mathbb{Z}_2 \rightarrow 1$ \\
				\hline
			\end{tabular}
		\end{center}
		\caption{Summary of unwinding 1+1D bosonic SPT phases by symmetry extension studied in Sec.~\ref{sec:bosonic}. Note that
			although $H^2(\mathbb{D}_8,U(1))=\Z_2$ has a nontrivial SPT class, still the SPT state in $H^2({\mathbb{Z}_2^2},U(1))=\Z_2$ can be deformed to a trivial product state in an extended $\mathbb{D}_8$-symmetry. 		\label{table:table5}}
	\end{table}
}

\newpage
\subsection{More remarks}

We conclude by providing more remarks on related physics and other works appeared in the literature.

\begin{enumerate}
	\item 
	\emph{Relation to some recent works}:
	Ref.~\cite{WangWenWitten_PhysRevX.8.031048_symmetric} provides a generic description for
	unwinding bosonic SPT states protected by a finite group $G$ (for both unitary or anti-unitary, such as time reversal symmetry).
	In \cite{WangWenWitten_PhysRevX.8.031048_symmetric},
	it has been shown that
	when the dimensions of spacetime $d+1$ is larger or equal to 1+1D,
	given a cohomology group $H^{d+1}(G,U(1))$ and the consequential SPT state protected by $G$-symmetry,
	we can always find an appropriate finite group $K$ extension to trivialize the $\omega_{d+1}(g)=H^{d+1}(G,U(1))$
	by viewing it (i.e. pulling it back) in a larger $\tilde G$ via a suitable
	$1  \longrightarrow {K}  \overset{i}{\longrightarrow} \tilde{G} \overset{s}{\longrightarrow}  G \longrightarrow 1$
	(See more details in \cite{WangWenWitten_PhysRevX.8.031048_symmetric}).
	Ref. \cite{WangWenWitten_PhysRevX.8.031048_symmetric} also
	provides the physical meaning for the above successful group extension
	in terms of three kinds of topological boundary/interface constructions:
	(i) $\tilde{G}$-symmetry extended boundary of $G$-SPT state: all the groups ($K, \tilde{G}$ and $G$) are symmetry groups and ungauged.
	(ii) ${G}$-symmetric $K$-gauged boundary  of $G$-SPT state: Only $K$ is dynamically gauged out of the total $ \tilde{G}$.
	(iii) $\tilde{G}$-gauged boundary of $G$-gauge theory: all the groups ($K, \tilde{G}$ and $G$) are dynamically gauged.
	A more recent work Ref.~\cite{1801.05416}
	explores the relations between the symmetry-breaking and symmetry-extension constructions, especially after gauging the bulk of group $G$.
	Ref.~\cite{1706.09782HuWanWu} provides the symmetry-breaking construction (breaking $G$ to $G'$) for topological order states in 2+1D.
	Another work Ref.~\cite{1712.09542.YT}
	also provides a very helpful exploration with certain mathematical rigor
	on the corresponding ``anomaly'' related to $H^{d+1}(G,U(1))$
	and  $H^{3}(G,K)$, after gauging the finite group $K$.

	\item 
	\emph{Unwinding fermionic SPT states v.s. Trivializing the topological terms from cobordism groups}:
	In contrast to the works of Refs.\cite{WangWenWitten_PhysRevX.8.031048_symmetric, 1801.05416, 1712.09542.YT} 
	mostly focusing on bosonic states,
	our work has implemented the general ideas to fermionic SPT states with short-range entanglement.
	The fermionic SPT states we studied (in Table.~\ref{table:web}) can also be regarded as topological invariants 
	generated from cobordism group calculations.
	Their precise cobordism groups can be found in  Refs.\cite{KapustinThorngrenWang_2015JHEP_Cobordism, 1711.11587.GPW}.
	Therefore, we may interpret our ``unwinding fermionic SPT states'' as
	the mathematical equivalent statement to trivialize the
	topological terms from cobordism group $\Omega^{d+1,\text{Spin/Pin}^{\pm}}_{\text{tors}}(BG,U(1))$
	where $BG$ means the classifying space of $G$ and tors means the torsion part,
	by lifting it (pulling it back) to the corresponding $\tilde{G}$'s cobordism group. 
	
	\item 
	\emph{Non-perturbative global anomaly, the finite torsion group in classifications}:
	As noted in Ref.\cite{WangWenWitten_PhysRevX.8.031048_symmetric},
	the SPT unwinding state procedure \emph{only} works for 
	SPT states obtained from a finite group (say, $\Z_n$, the torsion part) in the SPT classifications.
	The SPT unwinding state procedure does not work for the free part $\Z$ in the topological phase classifications.
	The finite group $\Z_n$ corresponds to non-perturbative global anomalies on the boundary of SPT state
	that can be trivialized by suitable group extensions.
	Instead, the free part $\Z$ corresponds to perturbative anomalies on the boundary of SPT state
	that \emph{cannot} be trivialized by any finite group extension.
	
	\item 
	\emph{General statements and proofs}:
	We provide the proof of the existence of symmetry-extension
	for 1+1D bosonic SPT systems with finite group symmetries based on the properties of Schur cover in Sec.~\ref{sec:Schur}.
	This can be viewed as the special case for the proof of \cite{WangWenWitten_PhysRevX.8.031048_symmetric} (for 1+1D and above dimensions)
	and the proof recently given in \cite{1712.09542.YT}.
	
	\item \emph{Some connections to quantum information processing}:
	The approach that we have used to unwind the SPT phases relies on (i) supplying generalized singlets that are invariant under the extended symmetry and (ii) then applying a sequence of SWAP gates. The SWAP gates in all cases considered commute with the extended symmetry. It is also interesting to note that after the unwinding procedure, the original degrees of freedom become trivialized while the supplied singlets are returned. These singlets act like a catalysis for the unwinding. The only effect on the catalytic singlets is that they are moved by one lattice site. In quantum information theory, a similar phenomenon appears in the conversion of quantum states that are made possible by supplying certain entangled states, i.e., entanglement catalysis~\cite{Nielsen99,JonathanPlenio}. 
	
	SWAP gates are the essential operation in our unwinding procedure, but these gates do not create entanglement nor enable universal quantum computation. However, computation using certain class of gates, called matchgates~\cite{Valiant}, can be efficiently simulated by a classical computer, but it can be made quantum computationally universal by introducing SWAP gates into the set of allowable gates~\cite{JozsaMiyake}. Matchgate quantum computation can also be formulated in terms of Majorana fermions~\cite{TerhalDiVincenzo} and the generalization of the Kitaev chains to quantum error correction codes has also be studied~\cite{MajoranaCodes}. It will be interesting to explore the connection between the Majorana fermion codes and the fermionic SPT phases.

\end{enumerate}

\subsection{Unwinding short-range entanglement v.s. long-range entanglement, and a gravity theory}
	
\label{sec:unwind-SRE-LRE}
	
In our work, we had considered several concrete SPT examples and how to unwinding their short-range entanglements.
For 1+1D fermionic SPT states,
our approach on unwinding short-range entanglement only works for certain ``\emph{even}'' number of 1+1D Kitaev Majorana fermionic chains
or bosonic chains like the Haldane spin chain.
It is curious to notice that the recent work of Dijkgraaf and Witten.~\cite{2018arXiv180403275DWitten}
achieves lifting the 0+1D Majorana zero modes of a \emph{single} 1+1D Kitaev chain, 
by coupling the system to a 1+1D topological gravity theory.
Since the single Kitaev chain is protected by no symmetry (except of the $\ztwof$ fermion parity),
thus it is a long-range entangled state in the sense of Wen's definition \cite{1610.03911Wen, Wen_2013_PhysRevDSPTAnomaly}. 

Here let us briefly review the meanings of short-range entanglement (SRE) and long-range entanglement (LRE) in this context.
In 1+1D, most of quantum mechanical systems we studied in Table.~\ref{table:web} are SRE.
Most of bosonic/fermionic chains as SPT states become trivial when we removing the global symmetries.
However, 
we cannot remove (an odd layer of) Kitaev chain's entanglement structure by local unitary transformation, 
unless we break the $\ztwof$ (which necessarily breaks a fermionic system to a bosonic system).
Thus, a single Kitaev chain is the only known example that is LRE in 1+1D with
an invertible fermionic topological order, described by
an invertible spin TQFT at its low energy. It is robust against any local perturbation as long as we keep the $\ztwof$-fermion parity symmetry.
To recap, we list below some representative examples for comparison ---

SRE examples: 1+1D Haldane spin/bosonic chains. 1+1D even numbers of layers of Kitaev fermionic chains.
SPT states in other dimensions.
Any invertible TQFT (iTQFT) that has \emph{no} invertible topological order (iTO in \cite{1610.03911Wen}).

LRE examples: 
A 1+1D single layer Kitaev fermionic chain. A 2+1D integer quantum Hall state (including the filling fraction $\nu=1$). A 2+1D $E_8$ state.
Any example of invertible topological orders (iTO) and topological orders, etc.

In our understanding, we can interpret Dijkgraaf and Witten's way of lifting the Majorana zero mode \cite{2018arXiv180403275DWitten}
as the spontaneous breaking of $\ztwof$-fermion parity symmetry
\emph{only} on the 0+1D boundary, while  $\ztwof$ can be preserved in the 1+1D bulk.
By applying Dijkgraaf and Witten's idea \cite{2018arXiv180403275DWitten},
in the future,
we may be able to achieve the unwinding of the \emph{long-range entanglement} of a Kitaev's Majorana fermionic chain
by coupling it to another \emph{long-range entangled} gravity theory.

\section{Acknowledgments}

A.P. is grateful to J.P. Ang, N. Tantivasadakarn and the users of ``Physics Stack Exchange" and ``Physics Overflow" websites for helpful discussions.
J.W. thanks  Edward Witten for his feedback on interpreting Ref.~\cite{2018arXiv180403275DWitten}.
J.W. gratefully acknowledges the Corning Glass Works Foundation Fellowship and NSF Grant PHY-1314311 and PHY-1606531.
A.P. and T.-C.W acknowledge support from NSF via Grants No. PHY 1620252, No. PHY 1333903 and No. PHY 1314748.

\appendix

\section{Realizing fermionic SPT phases by stacking Kitaev chains}

\label{app:fermion_hamiltonian_stracking}
In this section, we present model Hamiltonians using layers of the so-called Kitaev Majorana chain, which realize short-range-entangled fermionic phases corresponding to the five Altland-Zirnbauer classes that have a non-trivial classification in the free limit in 1+1 d. These classes are D, DIII, BDI, AIII and CII. 
To connect with the classification in the presence of interactions, we consider particular global symmetries of  D, DIII, BDI, AIII and CII: 
$\ztwof$, $\mathcal{Z}_4^T$, $\ztwo^T \times \ztwof$,
$\mathcal{U}(1) \times \ztwo^T$, and 
$\frac{(\mathcal{U}(1) \rtimes \mathcal{Z}_4^C)}{\mathcal{Z}_2^f} \times \mathbb{Z}_2^T$ symmetries.

\subsection{Class D {($\ztwof$-symmetry)}}
\label{sec:D}
\begin{figure}[!htbp]
	\centering	
	\includegraphics[width=100mm]{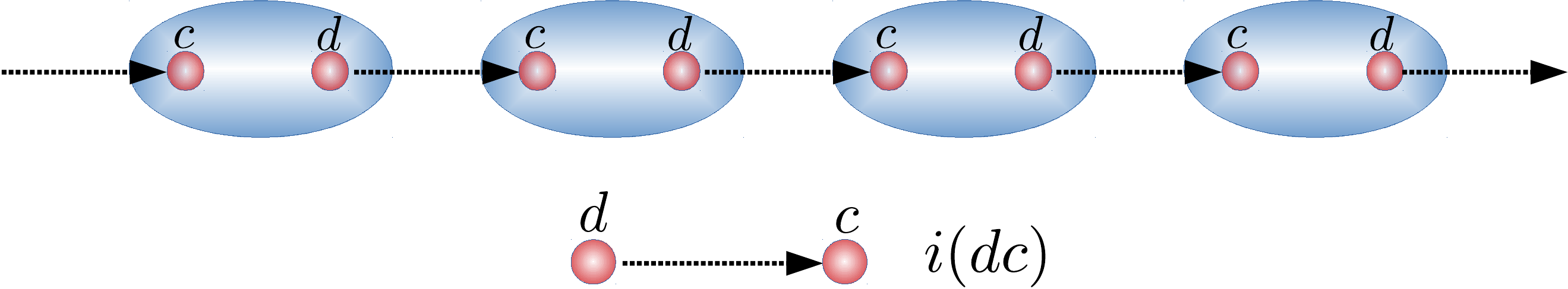}
	\caption{The Kitaev chain. \label{fig:Kitaev}}
\end{figure}

Let us start with the Hamiltonian for the Kitaev chain~\cite{Kitaev_2001PhysicsUspekhi_KitaevChain} which is a model of spinless fermions (on-site Hilbert space of a single fermionic mode) on a one-dimensional chain as shown in Fig.~\ref{fig:Kitaev}:
\begin{equation}\label{eq:Kitaev chain single layer}
H_{{\text{D}}} = i \sum_{k }  d_k c_{k+1}.
\end{equation}
$c_i$ and $d_i$ are Majorana operators which are defined in terms of creation and annihilation operators of the fermion mode, $\psi_i, \psi_i^\dagger$ as follows
\begin{equation}
c_i = \psi_i^\dagger + \psi_i, ~~~ d_i = i \left(\psi_i^\dagger - \psi_i  \right).
\end{equation}

\begin{figure}[!htbp]
	\centering	
	\includegraphics[width=100mm]{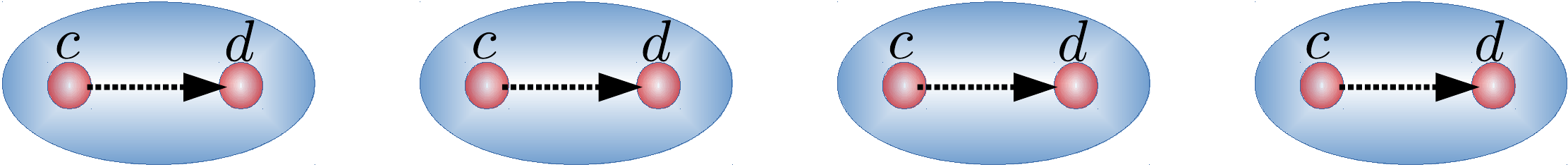}
	\caption{The trivial Majorana chain. \label{fig:Trivial}}
\end{figure}

If no other symmetries except $\ztwof$ is taken into consideration, this model of free fermions belongs to class D. SRE phases of this class have a $\ztwo$ classification in the non-interacting limit~\cite{SchnyderTyuFurusaki_2009AIPConference_PeriodicTable,Kitaev_2009AIPConference_PeriodicTable} and $H_D$ is a representative of the non-trivial phase. Since this phase is stable to interactions ~\cite{KapustinThorngrenWang_2015JHEP_Cobordism,FreedHopkin_2014arXiv_SREInvertibleField}, \emph{$H_D$ is a representative of a non-trivial phase of interacting fermions with no symmetries other than $\ztwof$}. For completeness, we also mention a representative of the trivial phase with the same symmetries shown in Fig.~\ref{fig:Trivial}.
\begin{eqnarray}\label{eq:Trivial Hamiltonian}
H^0_D = i \sum_{k }  c_k d_{k}.
\end{eqnarray}

\subsection{Class DIII {($\mathcal{Z}_4^T$-symmetry)}}~\label{sec:DIII}

\begin{figure}[!htbp]
	\centering	
	\includegraphics[width=100mm]{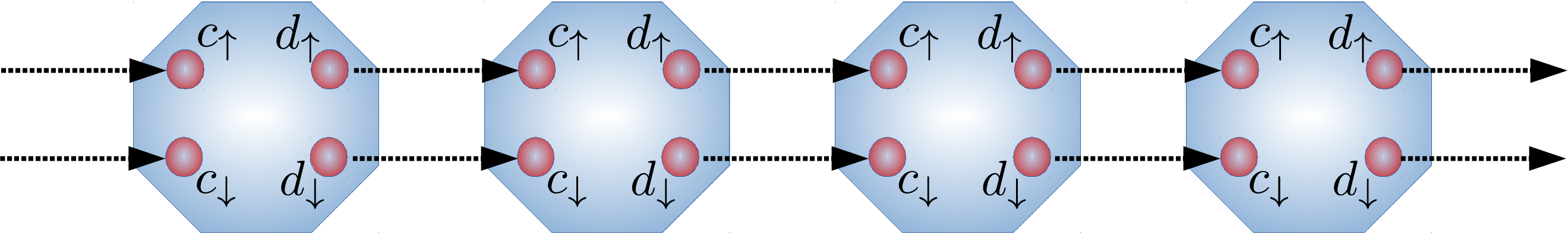}
	\caption{Non-trivial DIII chain. \label{fig:DIII}}
\end{figure}

We now consider a Hamiltonian with two species of fermions per unit site, which we will label as $\uparrow$ and $\downarrow$, constructed using two layers of Kitaev chains as shown in fig.~\ref{fig:DIII},
\begin{eqnarray}\label{eq:Kitaev chain two layers.}
H_{\text{DIII}} &=& i \sum_{k }  \sum_{\sigma = \uparrow, \downarrow} d_{\sigma,k}~c_{\sigma,k+1}.
\end{eqnarray}
This Hamiltonian commutes with the anti-unitary time-reversal operator $\mathcal{T}$ , 
\begin{eqnarray}
\mathcal{T}  &=& \prod_{k }  \exp{-\frac{\pi}{4} \left(c_{\uparrow} c_{\downarrow} + d_{\uparrow} d_{\downarrow}\right)_k} ~ \mathcal{K} = \prod_{k }  \frac{\left(1-c_{\uparrow} c_{\downarrow}\right)_k }{\sqrt{2}}  \frac{\left(1-d_{\uparrow} d_{\downarrow}\right)_k }{\sqrt{2}} ~ \mathcal{K}, \\
\mathcal{T}^2 &=& \prod_{k }  \left(i c_{\uparrow,k} d_{\uparrow,k}\right) \left(i c_{\downarrow,k} d_{\downarrow,k}\right) = \pf,
\end{eqnarray}
where $\pf$ is the fermion parity and $\mathcal{K}$ denotes complex conjugation, which has the following action 
\begin{eqnarray} \label{eq:complex conjugation}
\mathcal{K} i \mathcal{K} &=& -i, ~~~ \mathcal{K} c_\alpha \mathcal{K} = c_\alpha, ~~~ \mathcal{K} d_\alpha \mathcal{K} = -d_\alpha. 
\end{eqnarray}
We denote this symmetry group as $\mathcal{Z}_4^T$ and should be distinguished from $\mathbb{Z}_4^T$ defined in the previous subsection. The action of $\mathcal{T}$ can be seen in a more conventional form on creation and annihilation operators defined in the usual way.
\begin{eqnarray}
\psi_{\sigma,k} &=& \frac{1}{2}\left(c_{\sigma} + i d_{\sigma}\right)_k, ~~~ \psi_{\sigma,k}^\dagger =  \frac{1}{2}\left(c_{\sigma} - i d_{\sigma}\right)_k, \\
\mathcal{T} &=&  \prod_{k }  \left(\exp{-i \frac{\pi}{2}  \sigma^y_{\alpha \beta} \psi^\dagger_\alpha \psi_\beta}\right)_k \mathcal{K} ~=~ \prod_{k } \left(\exp{-i \pi {S}_y} \right)_k\mathcal{K}, \\
\mathcal{T} \psi_{\alpha,k} \mathcal{T}^{-1} &=& i \sigma^y_{\alpha,\beta}~ \psi_{\beta,k}.
\end{eqnarray}

With the symmetry $\mathcal{G} = \mathcal{Z}_4^T$, this free fermion model belongs to class DIII. SRE phases of this class has a $\ztwo$ classification in the non-interacting limit~\cite{SchnyderTyuFurusaki_2009AIPConference_PeriodicTable,Kitaev_2009AIPConference_PeriodicTable} and $H_{\text{DIII}}$ is a representative of the non-trivial phase. Since this phase is stable to $\mathcal{T}$ invariant interactions~\cite{KapustinThorngrenWang_2015JHEP_Cobordism,FreedHopkin_2014arXiv_SREInvertibleField}, \emph{$H_{\text{DIII}}$ is a representative of a non-trivial phase of interacting fermions with $\mathcal{G} = \mathcal{Z}_4^T$ symmetry}. For completion, we also mention a representative of the trivial phase with the same symmetries
\begin{eqnarray}
H^0_{\text{DIII}} = i \sum_{k }  \sum_{\sigma = \uparrow, \downarrow} c_{\sigma,k}~d_{\sigma,k},
\end{eqnarray}
which is simply two copies of the trivial Hamiltonian.~\ref{eq:Trivial Hamiltonian}.

\subsection{Class BDI {($\ztwo^T \times \ztwof$-symmetry)}}~\label{sec:BDI}
\begin{figure}[!htbp]
	\centering	
	\includegraphics[width=100mm]{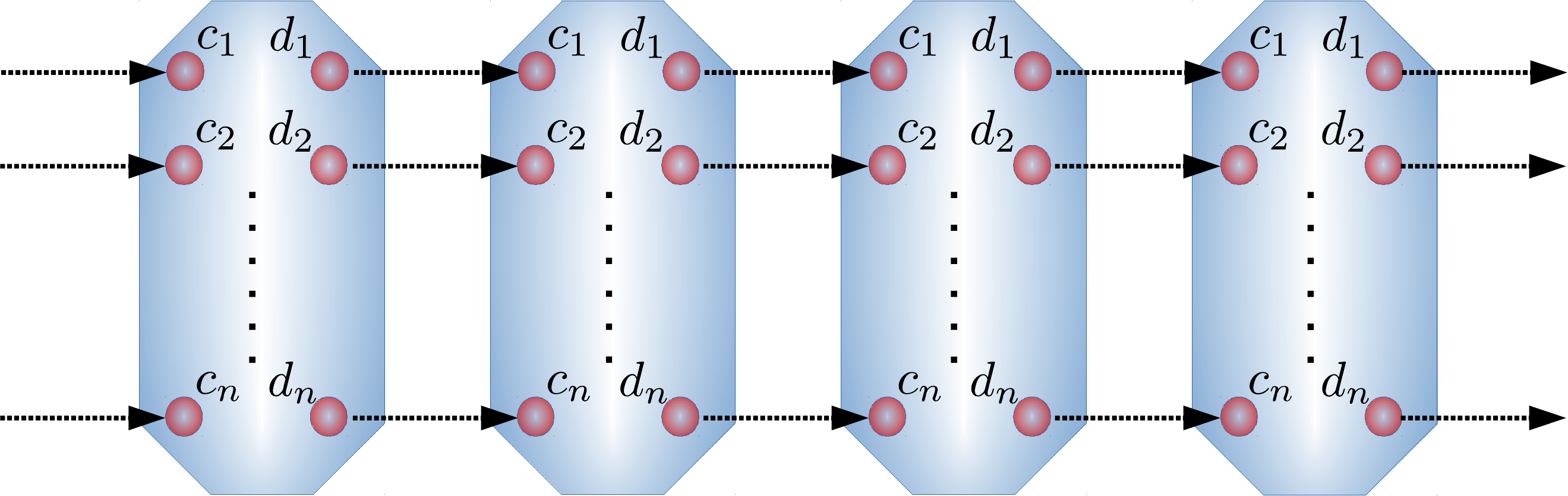}
	\caption{Stacked Kitaev chains. \label{fig:Kitaev_stack}}
\end{figure}

Let us once again consider the Kitaev chain Hamiltonian of Eq.~\ref{eq:Kitaev chain single layer}. It can be checked that the Hamiltonian is invariant under an anti-unitary time-reversal operation that only involves complex conjugation, $\mathcal{T} = \mathcal{K}$ which satisfies $\mathcal{T}^2 = 1$ and we call the group $\ztwo^T$. The full symmetry group is $\mathcal{G} = \ztwo^T \times \ztwof$. With this symmetry being considered, the free-fermion Kitaev Hamiltonian.~\ref{eq:Kitaev chain single layer} belongs to class BDI. SRE phases of this class has a $\mathbb{Z}$ classification in the non-interacting limit ~\cite{SchnyderTyuFurusaki_2009AIPConference_PeriodicTable,Kitaev_2009AIPConference_PeriodicTable}. We can think of the Kitaev chain to be a \emph{generating} Hamiltonian for all the non-trivial phases in this class by stacking as shown in Fig.~\ref{fig:Kitaev_stack}. Let us list representatives of each non-interacting phase labeled by $n \in \mathbb{Z}$:
\begin{eqnarray}
H^{(n)}_{\text{BDI}} &=& i \sum_{\alpha = 1}^{|n|} \sum_{k }  d_{\alpha,k}~c_{\alpha,k+1} ~~ \forall n \in \mathbb{Z}^+, \\
H^{(n)}_{\text{BDI}} &=& i \sum_{\alpha = 1}^{|n|} \sum_{k }  c_{\alpha,k}~d_{\alpha,k+1} ~~ \forall n \in \mathbb{Z}^- ,\\
H^{(0)}_{\text{BDI}} &=& i \sum_{k }  c_k d_{k}.
\end{eqnarray}

In the presence of interactions, it was shown in~\cite{FidkowskiKitaev_2010PhysRevB_InteractionBDI} that the $n=8$ Hamiltonian can be smoothly deformed to eight copies of $H^{(0)}_{\text{BDI}}$ without closing the gap. This means that in the presence of interactions, the SPT phases for this global symmetry has a $\mathbb{Z}_8$ classification whose representatives are $H^{(1)}_{\text{BDI}},\ldots,H^{(8)}_{\text{BDI}}$. 

\subsection{Class AIII {($\mathcal{U}(1) \times \ztwo^T$-symmetry)}}~\label{sec:AIII}
If we consider the even members of $H^{(n)}_{\text{BDI}}$, we can associate a $\mathcal{U}(1)$ symmetry in addition to time-reversal and commutes with it. Let us consider $H^{(2)}_{\text{BDI}}$
\begin{equation} \label{eq:Kitaev chain 2 layers}
H^{(2)}_{\text{BDI}} = i \sum_{\alpha=1}^{2} \sum_{k }  d_{\alpha,k} c_{\alpha,k+1},
\end{equation}
and the following $\mathcal{U}(1)$ operator which commutes with $\mathcal{T} = \mathcal{K}$,
\begin{eqnarray}
D(\theta) = \prod_{k }  \exp{-\frac{\theta}{2} \left(c_{1} c_{2} + d_{1} d_{2}\right)_k} .
\end{eqnarray}
To show invariance of Eq.~\ref{eq:Kitaev chain 2 layers} under $D(\theta)$, let us first look at the action on the Majorana operators,
\begin{eqnarray}
D(\theta) \begin{pmatrix}
c_1 \\
c_2
\end{pmatrix}_k D(\theta)^\dagger &=& \begin{pmatrix}
\cos \theta & \sin \theta \\
-\sin \theta & \cos \theta
\end{pmatrix}
\begin{pmatrix}
c_1 \\
c_2
\end{pmatrix}_k, \\
D(\theta) \begin{pmatrix}
d_1 \\
d_2
\end{pmatrix}_k D(\theta)^\dagger &=& \begin{pmatrix}
\cos \theta & \sin \theta \\
-\sin \theta & \cos \theta
\end{pmatrix}
\begin{pmatrix}
d_1 \\
d_2
\end{pmatrix}_k.
\end{eqnarray}
Now, we write the Hamiltonian.~\ref{eq:Kitaev chain 2 layers} in a suggestive form which makes invariance under $D(\theta)$ manifest,
\begin{eqnarray}
H^{(2)}_{\text{BDI}} &=& i \sum_{k }  \begin{pmatrix}
d_1  & d_2
\end{pmatrix}_k 
\begin{pmatrix}
c_1  \\ c_2
\end{pmatrix}_{k+1}, \\
D(\theta) H^{(2)}_{\text{BDI}} D(\theta)^\dagger &=& i \sum_{k }  \begin{pmatrix}
d_1  & d_2
\end{pmatrix}_k 
\begin{pmatrix}
\cos \theta & -\sin \theta \\
\sin \theta & \cos \theta
\end{pmatrix}
\begin{pmatrix}
\cos \theta & \sin \theta \\
-\sin \theta & \cos \theta
\end{pmatrix}
\begin{pmatrix}
c_1  \\ c_2
\end{pmatrix}_{k+1} \nonumber  \\&=& H^{(2)}_{\text{BDI}} .
\end{eqnarray}

Hence, the symmetry group is $\mathcal{G} = \mathcal{U}(1) \times \ztwo^T$. Note that $D(\pi) = \pf$ and hence we have used calligraphic script to denote the $\mathcal{U}(1)$ symmetry. This free model belongs to class AIII and SRE phases of this class has a $\mathbb{Z}$ classification. The representatives of each phase $n \in \mathbb{Z}$ can be obtained by considering the even members, $H^{{(2n)}}_{\text{BDI}}$. In the presence of interactions respecting $\mathcal{U}(1) \times \ztwo^T$ , the classification reduces to $\mathbb{Z}_4$ whose representatives are simply $H^{(2)}_{\text{BDI}},~H^{(4)}_{\text{BDI}},~H^{(6)}_{\text{BDI}},~H^{(8)}_{\text{BDI}}$.

\begin{figure}[!htbp]
	\centering	
	\includegraphics[width=100mm]{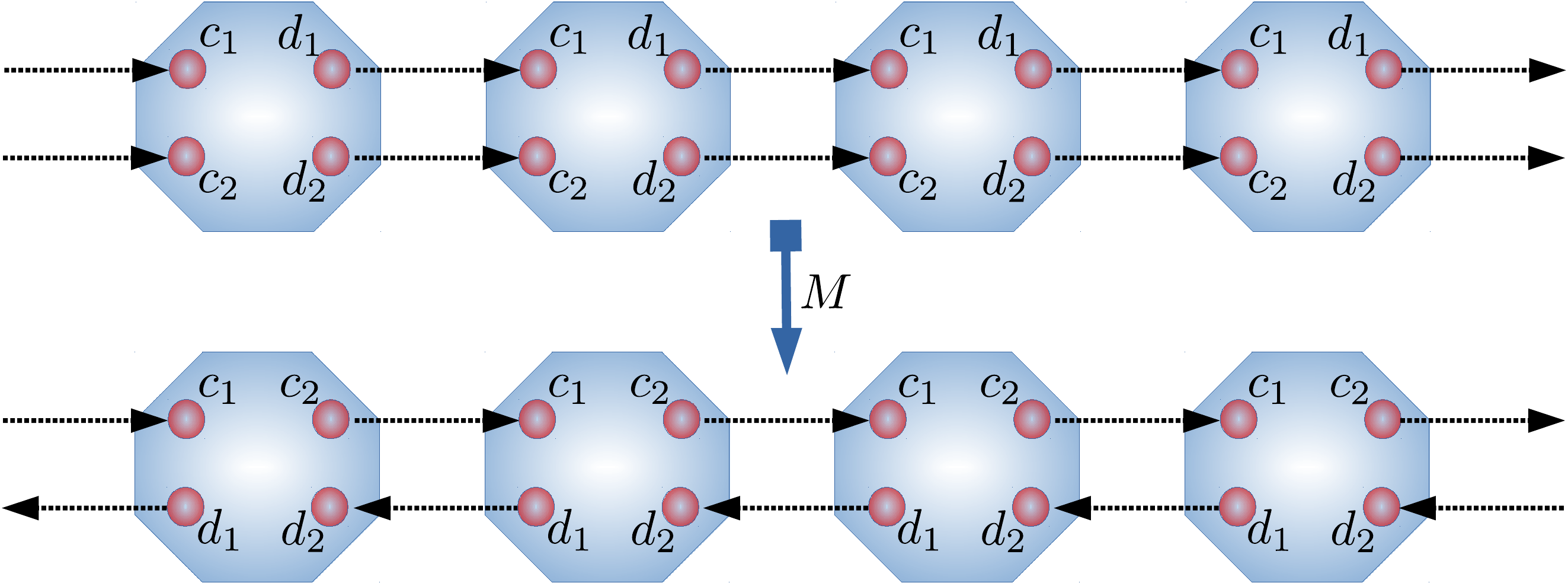}
	\caption{Non-trivial AIII chain before and after change of basis. \label{fig:AIII}}
\end{figure}

To make things clearer and for future convenience, we perform an on-site basis  change using the unitary operator, $M \equiv \prod_{k }  \exp{\frac{\pi}{4} \left(c_{2} d_{1} \right)_k}$ as shown in Fig.~\ref{fig:AIII}. Let us see the action on $H^{(2)}_{\text{BDI}}$:
\begin{eqnarray} 
H_{\text{AIII}} &\equiv& M H^{(2)}_{\text{BDI}} M^\dagger = i \sum_{k } \left(c_{2,k} c_{1,k+1} - d_{2,k} d_{1,k+1}\right),\label{eq:AIII Hamiltonian} \\
\mathcal{S} &\equiv& M \mathcal{T} M^\dagger  = M M^T \mathcal{K} = \prod_{k }  \left(c_2 d_{1}\right)_k \mathcal{K},\\
V(\theta)&\equiv& MD(\theta) M^\dagger = \prod_{k }  \exp{\frac{\theta}{2} \left(c_1 d_1 - c_2 d_2\right)_k}.
\end{eqnarray} 

Let us rewrite the new Hamiltonian $H_{\text{AIII}}$ in terms of the following fermion creation and annihilation operators,
\begin{eqnarray}
\psi_{1,k} &\equiv& \frac{1}{2} \left(c_1-id_1\right)_k,~~~\psi^\dagger_{1,k} = \frac{1}{2} \left(c_1+id_1\right)_k, \\
\psi_{2,k} &\equiv& \frac{1}{2} \left(c_2+id_2\right)_k,~~~\psi^\dagger_{2,k} = \frac{1}{2} \left(c_2-id_2\right)_k, \\
H_{\text{AIII}} &=& 2i \sum_{k }  \left(\psi^\dagger_{2,i} \psi_{1,i+1} + \psi_{2,i} \psi^\dagger_{1,i+1}\right).
\end{eqnarray}
First, note that the $\mathcal{U}(1)$ represented by $V(\theta)$ is now manifest in this form of the Hamiltonian. 
If we interpret fermions labeled $1$ and $2$ to be residing on even and odd sites of a chain, $H_{\text{AIII}}$ can be viewed as the bipartite hopping model~\cite{GadeWegner_1991NuclPhysB_Bipartite,Gade_1993NuclPhysB_Bipartite,KaiTeoSchnyderRyu_RevModPhys2016_ClassificationOfTopWithSymm} $\sum_{m,n} t_{mn} \psi^\dagger_m \psi_n$ with $t_{mn} = t^*_{nm}$ and has the following chiral symmetry:
\begin{eqnarray}
\mathcal{S} \psi_m \mathcal{S}^{-1} &=& (-1)^m \psi^\dagger_m, \\
\mathcal{S} i \mathcal{S}^{-1} &=& -i.
\end{eqnarray} 
For clarity, let us write down the Hamiltonian representatives and the symmetry operators of the four SRE phases written in the new form, labeled $n = 1,2,3,4$.
\begin{eqnarray}
H^{(n)}_{\text{AIII}} &=& i \sum_{\alpha=1}^{n} \sum_{k }  \left(c_{\alpha,2,k} c_{\alpha,1,k+1} - d_{\alpha,2,k} d_{\alpha,1,k+1}\right),\\
\mathcal{S} &=& \prod_{k }  \prod_{\alpha= 1}^{n} \left(c_{\alpha,2} d_{\alpha,1}\right)_k \mathcal{K},\\
V(\theta)&=&  \prod_{k }  \exp{\frac{\theta}{2} \sum_{\alpha=1}^{n}\left(c_{\alpha, 1} d_{\alpha, 1} - c_{\alpha, 2} d_{\alpha, 2} \right)_k}.
\end{eqnarray} 

\subsection{Class CII ($\frac{(\mathcal{U}(1) \rtimes \mathcal{Z}_4^C)}{\mathcal{Z}_2^f} \times \mathbb{Z}_2^T$-symmetry)}
\begin{figure}[!htbp]
	\centering	
	\includegraphics[width=100mm]{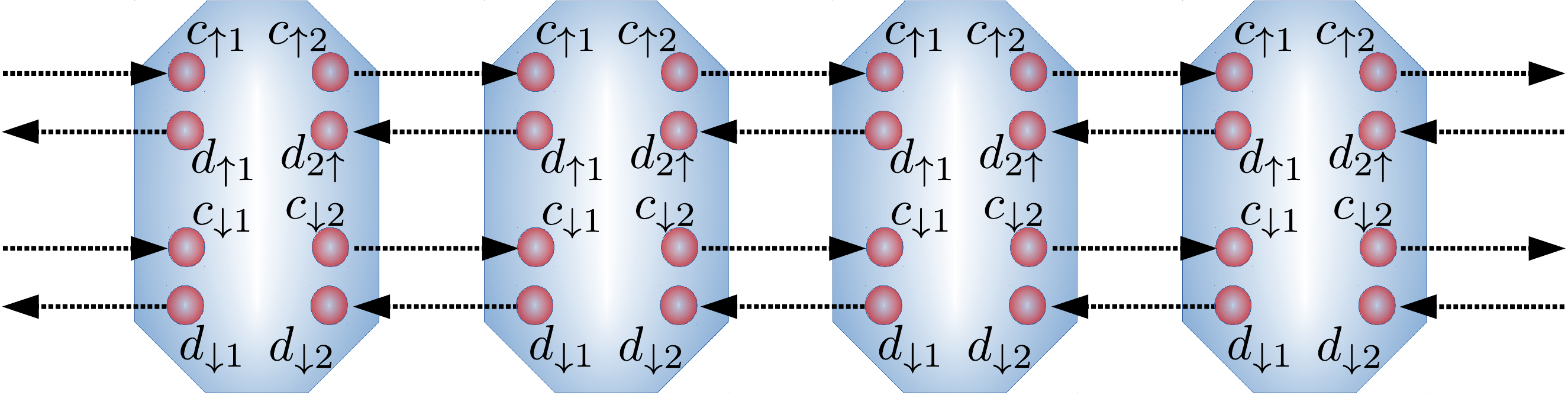}
	\caption{Non-trivial CII chain. \label{fig:CII}}
\end{figure}

Let us consider two layers of $H_{\text{AIII}}$~(\ref{eq:AIII Hamiltonian}) and label them as $\uparrow$ and $\downarrow$ as shown in Fig.~\ref{fig:CII},
\begin{equation} ~\label{eq:CII Hamiltonian}
H^{(2)}_{\text{AIII}} \equiv H_{\text{CII}} =   i  \sum_{k }  \sum_{\sigma=\uparrow, \downarrow} \left(c_{\sigma,2,k} c_{\sigma,1,k+1} - d_{\sigma,2,k} d_{\sigma,1,k+1}\right).
\end{equation}

Note that this contains four fermion species per unit cell labeled by $a = 1,2$ and $\sigma = \uparrow, \downarrow$. We can now define a unitary charge conjugation symmetry that commutes with $H_{\text{CII}}$ as follows,
\begin{equation}
\mathcal{C} = \prod_{k }  \exp{\frac{\pi}{4} \sum_{a=1}^2 \left(c_{\downarrow,a} c_{\uparrow,a}-d_{\downarrow,a} d_{\uparrow,a}\right)_k}.
\end{equation}
The action of $\mathcal{C}$ is best viewed on the creation and annihilation operators defined previously:
\begin{eqnarray}
\psi_{\sigma,1,k} &=& \frac{1}{2} \left(c_{\sigma,1}-i d_{\sigma,1}\right)_k,~~~\psi^\dagger_{\sigma,1,k} = \frac{1}{2} \left(c_{\sigma,1}+id_{\sigma,1}\right)_k, \\
\psi_{\sigma,2,k} &=& \frac{1}{2} \left(c_{\sigma,2}+i d_{\sigma,2}\right)_k,~~~\psi^\dagger_{\sigma,2,k} = \frac{1}{2} \left(c_{\sigma,2}-id_{\sigma,2}\right)_k, \\
\mathcal{C} \psi_{a,\alpha,k} \mathcal{C}^{-1} &=& i \sigma^y_{\alpha, \beta} \psi^\dagger_{a,\beta,k}.
\end{eqnarray}
Note that $\mathcal{C}^2 = \pf$ and the group generated by it is $\mathcal{Z}_4^C$. Furthermore, $\mathcal{C}$ commutes with the chiral symmetry $\mathcal{S}$ but not with the $\mathcal{U}(1)$ symmetries making the symmetry group $\mathcal{G} =$ 
{$\frac{(\mathcal{U}(1) \rtimes \mathcal{Z}_4^C)}{\mathcal{Z}_2^f} \times \mathbb{Z}_2^T$}
\begin{eqnarray}
\mathcal{S}  &=& \prod_{k }  \prod_{\sigma= \uparrow,\downarrow}^{n} \left(c_{\sigma,2} d_{\sigma,1}\right)_k \mathcal{K},~~~~~ \mathcal{C}~\mathcal{S}~\mathcal{C}^{-1}=\mathcal{S},\\
V(\theta)&=&  \prod_{k }  \exp{\frac{\theta}{2} \sum_{\sigma=\uparrow,\downarrow}^{n}\left(c_{\sigma, 1} d_{\sigma, 1} - c_{\sigma, 2} d_{\sigma, 2} \right)_k},~~~~~\mathcal{C}~ V(\theta) ~\mathcal{C}^{-1}= V(-\theta).
\end{eqnarray}
With this symmetry, the free fermion Hamiltonian.~\ref{eq:CII Hamiltonian} belongs to class CII. SRE phases of this class has a $\mathbb{Z}$ classification in the non-interacting limit and $H_{\text{CII}}$ is the generating representative of the non-trivial phases via stacking in the manner described in the previous subsections. In the presence of symmetry respecting interactions however, the classification breaks down to $\ztwo$ and $H_{\text{CII}}$ is a representative of the non-trivial phase. Finally, for completion, let us also state the Hamiltonian that corresponds to the trivial phase for this symmetry group,
\begin{equation}
H^0_{\text{CII}} = i  \sum_{k }  \sum_{\sigma=\uparrow, \downarrow} \left(c_{\sigma,1,k} c_{\sigma,2,k} - d_{\sigma,1,k} d_{\sigma,2,k}\right).
\end{equation}

\bibliography{references}
\bibliographystyle{arXiv-new}

\end{document}